\documentclass[aps,prd,twocolumn,superscriptaddress,amsmath,twoside,amssymbepsfig,showpacs]{revtex4-1}
\usepackage{graphicx}
\usepackage{dcolumn}
\usepackage{bm}
\usepackage{natbib}
\usepackage{color}
\usepackage{epstopdf}
\usepackage[hyperindex,breaklinks]{hyperref}

\fontencoding{T1}

\newcommand{\araa}{{Annu.~Rev.~Astron.~Astrophys.}}
\newcommand{\aap}{{Astron.~Astrophys.}}
\newcommand{\apjl}{{Astrophys.~J.~Lett.}}
\newcommand{\pasp}{{Publ. Astron. Soc. Pacif.}}
\newcommand{\apjs}{{Astrophys.~J.~Supp.}}
\newcommand{\aj}{{Astron.~J.}}
\newcommand{\mnras}{{Mon. Not. R. Astron. Soc.}}
\newcommand{\jcap}{{J.~Cosmol. Astropart. Phys.}}
\newcommand{\physrep}{{Phys.~Rep.}}

\newcommand{\be}{\begin{equation}}
\newcommand{\ee}{\end{equation}}
\newcommand{\ba}{\begin{eqnarray}}
\newcommand{\ea}{\end{eqnarray}}

\newcommand{\LCDM}{$ \Lambda $CDM}

\newcommand{\fnl}{f_{\mathrm{NL}}}
\newcommand{\gnl}{g_{\mathrm{NL}}}
\newcommand{\anl}{a_{\mathrm{NL}}}
\newcommand{\nfnl}{n_{f_{\mathrm{NL}}}}
\newcommand{\nn}{\nonumber}

\begin{document}

\title{Improved Primordial Non-Gaussianity Constraints from Measurements\\of Galaxy Clustering and the Integrated Sachs-Wolfe Effect} 

\author {Tommaso Giannantonio}
\email{tommaso.giannantonio@usm.lmu.de}
\affiliation{Ludwig-Maximilians-Universit\"at M\"unchen, Universit\"ats-Sternwarte M\"unchen, Scheinerstr. 1, D-81679 M\"unchen, Germany}
\affiliation{Excellence Cluster Universe, Technical University Munich, Boltzmannstr. 2, D-85748 Garching, Germany}

\author {Ashley J. Ross}
\affiliation {Institute of Cosmology and Gravitation, University
of Portsmouth, Portsmouth, PO1 3FX, UK}

\author {Will J. Percival}
\affiliation {Institute of Cosmology and Gravitation, University
of Portsmouth, Portsmouth, PO1 3FX, UK}

\author {Robert Crittenden}
\affiliation {Institute of Cosmology and Gravitation, University
of Portsmouth, Portsmouth, PO1 3FX, UK}

\author {David Bacher}
\affiliation{Ludwig-Maximilians-Universit\"at M\"unchen, Universit\"ats-Sternwarte M\"unchen, Scheinerstr. 1, D-81679 M\"unchen, Germany}
\affiliation{Excellence Cluster Universe, Technical University Munich, Boltzmannstr. 2, D-85748 Garching, Germany}

\author {Martin Kilbinger}
\affiliation{Laboratoire AIM, CEA Saclay, Irfu/SAp (Service d'Astrophysique), Orme des Merisiers, B\^at 709, F-91191 Gif-sur-Yvette, France}

\author {Robert Nichol}
\affiliation {Institute of Cosmology and Gravitation, University
of Portsmouth, Portsmouth, PO1 3FX, UK}

\author {Jochen Weller}
\affiliation{Ludwig-Maximilians-Universit\"at M\"unchen, Universit\"ats-Sternwarte M\"unchen, Scheinerstr. 1, D-81679 M\"unchen, Germany}
\affiliation{Excellence Cluster Universe, Technical University Munich, Boltzmannstr. 2, D-85748 Garching, Germany}
\affiliation{Max-Planck-Institut f\"ur Extraterrestrische Physik, Giessenbachstr., D-85748 Garching, Germany}

\begin {abstract}
We present the strongest robust constraints on primordial non-Gaussianity (PNG) from currently available galaxy surveys,  combining large-scale clustering measurements and their cross-correlations with the cosmic microwave background. We update the data sets used by Giannantonio \emph{et al.} (2012), and broaden that analysis to include the full set of two-point correlation functions between all surveys. In order to obtain the most reliable constraints on PNG, we advocate the use of the cross-correlations between the catalogs as a robust estimator and we perform an extended  analysis of the possible systematics to reduce their impact on the results.  To minimize the impact of stellar contamination in our luminous red galaxy (LRG) sample, we use the recent Baryon Oscillations Spectroscopic Survey catalog of Ross \emph{et al.} (2011).  We also find evidence for a new systematic in the NVSS radio galaxy survey similar to, but smaller than, the known declination-dependent issue; this is difficult to remove without affecting the inferred PNG signal, and thus we do not include the NVSS auto-correlation function in our analyses.
 We find no evidence of primordial non-Gaussianity; for the local-type configuration we obtain for the skewness parameter $ -36 < \fnl < 45 $ at 95\% c.l. ($5 \pm 21$ at $1\sigma$) when using the most conservative part of our data set, improving previous results; we also find no evidence for significant kurtosis, parameterized by $\gnl$. In addition to PNG, we simultaneously constrain dark energy and find that it is required with a form consistent with a cosmological constant.
 
\end {abstract}

\pacs {}

\maketitle

\section {Introduction} \label {sec:intro}

It is widely assumed that the density perturbations that seeded the cosmic microwave background (CMB) anisotropies and later induced the formation of cosmic structure were produced by primordial quantum fluctuations in the early universe \citep{Lyth:2009zz}. An early inflationary phase is then necessary to stretch such fluctuations beyond the horizon, and to address a number of other issues related to the initial conditions of the big bang. While the general paradigm of inflation is widely accepted, the precise details of the physical mechanism that produced an inflationary phase in the early universe remain largely unconstrained (see Ref.~\cite{Byrnes:2010a} for a recent review). However, any specific inflationary model is testable, as it leaves observable traces in the later evolution. For example, the simplest single-field slow-roll inflation model predicts a nearly flat universe and near scale-invariance of the primordial perturbations, both of which have been verified e.g. by the WMAP satellite \cite{Larson:2011a,Komatsu:2011a,2012arXiv1212.5226H,2012arXiv1212.5225B}. Additionally, the primordial perturbations are predicted to be purely adiabatic, which is also indicated (see Ref.~\cite{Valiviita:2009a} for recent constraints on isocurvature). Finally, the distribution of primordial perturbations should be nearly Gaussian \citep{2003JHEP...05..013M,2003NuPhB.667..119A}: testing this last prediction is our present focus.

Determining the amount of primordial non-Gaussianity (PNG) is  instrumental in distinguishing between models of the early universe \citep{2010JCAP...12..030S}. This has been typically constrained from the higher-order statistics (such as the bispectrum) of the CMB \cite{Komatsu2010a,Komatsu:2011a}; the latest CMB measurements from the nine-year analysis of data from the WMAP satellite constrain local PNG, as quantified by the skewness parameter $\fnl$ (defined in Section~\ref{sec:theory}) to be $-3 < \fnl < 77$ at 95\% confidence \citep{2012arXiv1212.5226H,2012arXiv1212.5225B} \footnote{Since this paper was submitted, Planck results have been released, showing no significant PNG detections at a level of $ -8.9 < \fnl < 14.3 $ (95 \% c.l.) \cite{2013arXiv1303.5084P}.}. Recent CMB constraints on the kurtosis parameter $\gnl$ are $-7.4 \cdot 10^5 < \gnl < 8.2 \cdot 10^5$ \cite{2010PhRvD..81l3007S}.
Higher-order statistics of the large-scale structure (LSS) may also be used \cite{2010AdAst2010E..73L,2012MNRAS.425.2903S}, but the difficulty of this approach is in disentangling PNG from late-time non-Gaussianity produced by non-linear structure formation. This is also an issue if using the effect on the non-linear contributions to the  matter power spectrum \cite{2008PhRvD..78l3534T}, probed e.g. by weak gravitational lensing \cite{2012MNRAS.426.2870H}.
The abundance of massive clusters is also a sensitive, complementary probe of PNG \cite{Matarrese2000,Loverde2008,2010MNRAS.402..191P,2012JCAP...02..002A}, and is less sensitive to later non-linear evolution.

In addition to these established methods, a new technique
has been discovered and developed in the recent years: it has been shown that PNG of the local and orthogonal types alters the biasing of dark matter halos and galaxies with respect to the underlying density field, making the bias strongly scale-dependent; thus by measuring the galaxy bias it is possible to constrain PNG \cite{Dalal:2008a,2008ApJ...677L..77M,Valageas:2010a,2009MNRAS.396...85D,2010CQGra..27l4011D,2010AdAst2010E..89D,Giannantonio:2010a,2010PhRvD..82j3529D,2011PhRvD..84f3512D,2011PhRvD..84f1301D,2011PhRvD..84h3509H}. This has been exploited by some authors \cite{Slosar:2008a,2008PhRvD..78l3507A,Xia:2010a,Xia:2010b,Xia:2011a,Ross13fnl}, finding  constraints which are competitive with CMB bispectrum analyses: Ref.~\cite{Slosar:2008a} found $-29 < \fnl < 70$ at 95\%, while Ref.~\cite{Xia:2011a} reported hints of detection at $ 8 < \fnl < 88 $ (95\%) for the local model. A measurement of $\gnl$ was obtained from LSS data by Ref.~\cite{2010PhRvD..81b3006D} assuming $\fnl = 0$ by rescaling the $\fnl$ constraints by Ref.~\cite{Slosar:2008a}: $ -3.5 \cdot 10^5 < \gnl < 8.2 \cdot 10^5$ (95\%).
The bias of galaxy clusters is similarly affected by PNG, and has been used in combination with cluster abundances to constrain $\fnl$ \cite{2013arXiv1303.0287M}.
 The scale-dependent bias technique will remain competitive with the arrival of future galaxy and cluster surveys, and is potentially more powerful than the cosmic-variance limit of the CMB \cite{2009MNRAS.397.1125F,2010JCAP...07..020C,Giannantonio:2011a, 2012JCAP...04..005H,2012MNRAS.422...44P,2011MNRAS.414.1545F,2010PhRvD..82b3004C}.

The large-scale divergence in galaxy bias produced by PNG can also be observed through cross-correlations of galaxy surveys with the CMB, provided some of the CMB anisotropies are created locally.  This is conveniently provided by the integrated Sachs-Wolfe (ISW) effect \cite{Sachs:1967a}, where small secondary CMB anisotropies are induced by the decay of the gravitational potentials when the Universe undergoes a transition from a dark matter to a dark energy or curvature dominated phase. 
While the nature of dark energy is still unknown, a broad range of observations have confirmed its presence \cite{Frieman:2008a}, including observations of the ISW effect \citep[see Ref.][and references therein]{2012MNRAS.426.2581G}. In most realistic models the ISW effect is too small to be directly observed in the primary CMB temperature power spectrum, but it can be detected by cross-correlating the CMB temperature anisotropies with tracers of the potentials, such as galaxy catalogs \cite{Crittenden:1996a}. The latest compilations which examine multiple catalogs are those by Refs.~\cite{Ho:2008a} and \cite{Giannantonio:2008a} (G08 herein); the latter was recently updated in Ref.~\cite{2012MNRAS.426.2581G} (G12 herein) and we refer to this paper for a more detailed introduction and a longer description of the ISW effect and the data. These cross-correlations are sensitive to the galaxy bias, and therefore also depend on PNG as above.

In this paper we use an updated compilation of LSS  and ISW data to make robust and accurate PNG measurements. We first review the galaxy samples that we consider, revising those used by G12. In particular, we include the latest available luminous red galaxy (LRG) photometric redshift sample \cite{Ross:2011a}, also used for ISW measurements by Ref.~\cite{HM13}, which were calibrated using data from the Baryon Oscillation Spectroscopic Survey (BOSS) \cite{2013AJ....145...10D} CMASS sample. Refs.~\cite{Ross:2011a,Ho12,Ross12} have shown that properly addressing the possibility of spurious fluctuations in the observed number of galaxies is crucial in order to measure and interpret LSS measurements. Refs.~\cite{Ross13fnl,Huterer_calib,Pullen_qso} have shown that these issues are particularly important when measuring PNG. We address these concerns in all of our analyses and  study the impact of some well-known systematic effects present in the radio-galaxies of the NVSS survey.

We then explore the cosmological consequences of our data, using a nested-sampling Monte Carlo Markov Chain (MCMC) method and marginalizing over extra nuisance parameters to account for a range of possible uncertainties and systematics. We derive cosmological constraints for both PNG and dark energy, extending the data set to include both the galaxy-CMB and all the galaxy-galaxy correlations, and include their full covariance. Indeed, we demonstrate the importance of the cross-correlations between different galaxy catalogs as a robust method to estimate the amount of PNG in the data. 

The plan of this paper is as follows: we briefly review the theoretical basis of the LSS in the presence of PNG and the ISW effect in Section~\ref{sec:theory}. We then describe our data set and the most important systematics in Section~\ref{sec:data}, while Section~\ref{sec:method} clarifies how we model the observations. In Section~\ref{sec:results} we  discuss the cosmological interpretation of our data, before concluding in Section~\ref{sec:conclusion}.

\section {Theory} \label{sec:theory}

\subsection{Primordial Non-Gaussianity and the LSS}

We consider here  deviations from Gaussianity  at the three- and four-point level (skewness and kurtosis). For PNG of the local type, the Bardeen potential at early times ($z_{\star}$) is assumed to be written in terms of an auxiliary Gaussian potential $\varphi$ as
\ba \label{eq:fnl101}
\Phi (\mathbf{x}, z_{\star}) &=& \varphi(\mathbf{x}, z_{\star}) + \fnl \, \left[ \varphi^2(\mathbf{x}, z_{\star}) - \langle \varphi^2 \rangle(z_{\star}) \right] \nonumber \\
&~& + \gnl \left[ \varphi^3(\mathbf{x}, z_{\star}) - 3 \langle \varphi^2 \rangle (z_{\star}) \, \varphi(\mathbf{x}, z_{\star}) \right] \, ,
\ea
where $\fnl$ and $\gnl$ quantify the amount of PNG. In this work we will also study simple deviations from the local case, as described below.

It was originally shown by Ref.~\cite{Dalal:2008a} that primordial non-Gaussianity induces a strongly scale-dependent bias, whose features appear on the largest scales for both dark-matter halos and galaxies, and their analysis was later clarified and improved \cite{Matarrese:2008a,Slosar:2008a,Afshordi:2008a,Valageas:2010a,Giannantonio:2010a,Schmidt:2010a,Desjacques:2011a,2012JCAP...03..032S}. The scale dependence originates through the coupling of long- and short-wavelength modes in the density perturbations which arise in the presence of PNG by applying the peak-background split technique to Eq.~(\ref{eq:fnl101}). In Fourier space, we can use the  linear Poisson equation to show  that the late-time density can be related to the initial potential field by 
\be \tilde \delta(k,z) \simeq \alpha(k,z) \, \tilde \varphi (k, z_{\star}), \ee  where the tilde denotes Fourier transformation. Here, 
\be \label{eq:alpha}
\alpha(k,z) = \frac{2 \, k^2 \, T(k) \, D(z)}{3 \, \Omega_m \, H_0^2} \, \frac{g(0)}{g(z_{\star})} \, ,
\ee
where $T(k)$ is the density transfer function, $D(z)$ is the linear growth function, $g(z) \propto (1+z) \, D(z)$ is the potential growth function, for which $g(z_{\star}) / g(0) \simeq 1.306$.

In the presence of PNG, the Gaussian bias $b_{\mathrm{Gauss}}$ is altered by two corrective terms, giving an effective bias
\be \label{eq:deltab_all}
b_{\mathrm{eff}}(k, \fnl, \gnl) = b_{\mathrm{Gauss}} + \delta b (\fnl, \gnl) + \Delta b (k,\fnl,\gnl) \, .
\ee
The first, scale-independent correction $\delta b (\fnl, \gnl)$ is produced by the effect of PNG on the halo mass function and it is typically negligible compared with the scale-dependent part $\Delta b (k,\fnl,\gnl)$; in the following we will absorb it into the constant part of the bias, which we define as $b_1(\fnl,\gnl) \equiv b_{\mathrm{Gauss}} + \delta b (\fnl,\gnl) $. For local-type non-Gaussianity, the scale-dependent term is (at any given redshift) \cite{2012JCAP...03..032S}
\be
\Delta b^{\mathrm{loc}} (k, \fnl, \gnl) = \frac{\beta_f \, \fnl + \beta_g \, \gnl}{\alpha(k)} \, .
\ee
The coefficients $\beta_f$ and $\beta_g$ will in general depend on redshift and mass (or bias). Here the $\fnl$ part is
\be \label{eq:bfnl}
\beta_f = 2 \, \delta_c \, b_L \, \,
\ee
where  $\delta_c = 1.686$ is the spherical collapse threshold and $b_L \equiv b_1 - 1 $ is the Lagrangian bias. 

For the $\gnl$ contribution $\beta_g$, we use the fit to $N$-body simulations of the Edgeworth expansion introduced by Ref.~\cite{2012JCAP...03..032S}:
\ba \label{eq:gNL}\beta_g &\simeq& \kappa_3^{(1)}(M) \, \left[ -0.7 + 1.4 \, (\tilde \nu - 1)^2 + 0.6 \, \left( \tilde \nu - 1\right)^3 \right] \nonumber \\
&~& - \frac{d \kappa_3^{(1)} (M)}{d \ln \sigma^{-1}} \, \left( \frac{\tilde \nu - \tilde \nu^{-1}}{2} \right) \, ,
\ea
with
$\tilde \nu \equiv \sqrt{  \tilde \delta_c \, b_L + 1}$, 
where $\tilde \delta_c = 1.42 $, which was found by Ref.~\cite{2012JCAP...03..032S} to improve the agreement between the halo collapse model and $N$-body simulations with respect to the Press-Schechter value.  We adopt this value in the $\gnl$ fitting formulae to be consistent with the derivation of Ref.~\cite{2012JCAP...03..032S} (although we keep the standard value when fitting $\fnl$, as the two parts of the model are separable).
$\kappa_3^{(1)}(M)$ is the skewness of the density field for the case $\fnl=1$, smoothed at a mass $M$, and $\sigma$ the r.m.s. of the same field at the same scale. For the skewness and its derivative we use the fitting functions given in Eqs.~(46-47) of Ref.~\cite{2012JCAP...03..032S}.  
Though their redshift dependence is different, both $\fnl$ and $\gnl$ cause the bias to become scale-dependent with the same power of $k$, albeit with different amplitudes; this makes distinguishing between the two parameters challenging \cite{2012MNRAS.425L..81R}.

The fitting formula of Eq.~(\ref{eq:gNL}) comes with an important caveat, as it was obtained for a narrow range in halo mass and redshift \cite{2012JCAP...03..032S}; furthermore, it was shown that the effect of $\gnl$ changes depending on the halo occupation distribution of the used tracers, not only on the Gaussian bias, and this effect becomes more severe for less biased samples, such as in our case.
This formula is, however, the best currently available, and the form of the constraints obtained using it will be indicative of those that would be available using a more accurate fit. We therefore adopt the formula, testing for specific non-Gaussian deviations of this form, making sure that when inferences are drawn from the results we remember these might be different from true $\gnl$-like deviations.

Any residual theoretical uncertainties on the local $\fnl$ contribution to the bias are less significant, as the linear model of Eq.~(\ref{eq:bfnl}) has been tested both theoretically and numerically with $N$-body simulations by multiple authors \cite{Matarrese:2008a,Slosar:2008a,Afshordi:2008a,Valageas:2010a,Giannantonio:2010a,Schmidt:2010a,Desjacques:2011a,2012JCAP...03..032S}. Non-linear effects have also been extensively studied by some of these authors, and they were shown to be negligible on the scales of interest; finally, while general-relativistic effects are present on the largest scales, they are small and do not affect the PNG constraints from the bias \cite{2012PhRvD..86f3514Y}. Even if small residual inaccuracies were present, they would primarily affect the interpretation of a detection, while the probability of $\fnl$ being non-zero should not significantly change.

Scale-dependent bias was first used to constrain $\fnl$ from the LSS by Refs.~\cite{Slosar:2008a,Afshordi:2008a} with varying results, possibly caused by different assumptions for the bias evolution with redshift. Ref.~\cite{2010PhRvD..81b3006D} obtained bounds on $\gnl$ from the same data consistent with Gaussian initial conditions. Some later work \cite{Xia:2010a,Xia:2010b,Xia:2011a} found hints of detection of a positive $\fnl$, while more recent analyses that take into account possible systematic errors in the large-scale clustering measurements have found measurements consistent with no PNG \cite{Ross13fnl}. 

Similar expressions have been derived for other varieties of PNG, such as orthogonal and equilateral \cite{Schmidt:2010a,Desjacques:2011a}, featuring weaker or no scale dependence. Furthermore, some models also predict a scale dependence of $\fnl$ \cite{2005PhRvD..72l3518C,2010JCAP...02..034B}, often parameterized with a spectral index $\nfnl$: recently, the first constraint on this quantity has been found $\nfnl = 0.30^{+1.9}_{-1.2}$ at 95\% c.l. from CMB data \cite{2012PhRvL.109l1302B}.
Given the difficulty in distinguishing between different models with current bias measurements, and to be as general as possible, one can adopt a simpler phenomenological approach by allowing the scale-dependent term to have an arbitrary amplitude and scaling index. To keep the interpretation of the parameter constraints closer to the usual local case, we adopt a parameterization
\be \label{eq:anl}
\Delta b^{\mathrm{gen}}(k, \fnl) =   \frac{2 \fnl^{\mathrm{gen}} \, \delta_c \, b_L} {\alpha'(k, \anl)}  \, ,
\ee
where we have only one extra parameter $\anl$, changing the slope of the bias correction:
\be
\alpha'(k,\anl,z) = \frac{2 \, k^{\anl} \, T(k) \, D(z)}{3 \, \Omega_m \, H_0^2} \, \frac{g(0)}{g(z_{\star})} \, .
\ee
The local, scale-independent case is recovered for $\anl = 2$. 
Notice also that this can be related to the local spectral index as 
\be \label{eq:an}
k^{\anl} = k^{2} \, \left(\frac{k_{\mathrm{piv}}}{k}\right)^{\nfnl} \, ,
\ee
where $k_{\mathrm{piv}}$ is the pivot scale. We will use $k_{\mathrm{piv}} = 0.002 \, h$ Mpc$^{-1}$ in the following as in Ref.~\cite{Giannantonio:2011a}.

\subsection{The ISW Effect and Large-Scale Structure} 
For a recent, detailed summary of the ISW theory and its observational status, we refer to Ref.~\cite{2012MNRAS.426.2581G}. Briefly, if the gravitational potentials $\Phi, \Psi$ are evolving in time, they will imprint secondary temperature anisotropies $\Theta \equiv \delta T/T$ in the CMB in a direction $\hat{\mathbf{n}}$ given by
\be \label{eq:ISWT}
\Theta (\hat {\mathbf{n}}) = - \int e^{- \tau (\eta)} \, \left( \dot \Phi + \dot \Psi \right) [\eta, \hat {\mathbf {n}}(\eta_0 - \eta)] \, d \eta \, ,
\ee
where $\eta$ is conformal time, the dots are derivatives with respect to it, and $\tau$ is the optical depth of CMB photons.
Similarly, for a galaxy catalog $i$ with redshift distribution $\varphi_i(z) \equiv {dN_i}/{dz} (z)$, its expected overdensity is
\be
\delta_{g_i}(\hat {\mathbf {n}}) = \int b_{g_i}(z) \, \varphi_i (z) \, \delta (\hat {\mathbf {n}}, z) \, dz \, ,
\ee
where $b_{g_i} (z)$ is the galactic bias and $\delta (\hat {\mathbf {n}}, z) $ the dark matter overdensity.

From these relations, the auto- and cross-spectra in harmonic space can be constructed assuming linear theory as
\ba
C_l^{g_i g_j} &=& \frac{2} {\pi} \int dk \, k^2 \, P(k) \, W_l^{g_i}(k) \, W_l^{g_j}(k) \,  \nonumber \, \nonumber \\
C_l^{T g_i} &=& \frac{2} {\pi} \int dk \, k^2 \, P(k)  \, W_l^{T}(k) \, W_l^{g_i}(k) \, ;
\ea
here $P(k)$ is the linear matter power spectrum at $z = 0$, and 
 the sources for galaxies and the ISW temperature anisotropies are (assuming $\Phi = \Psi$)
\ba
W_l^T(k) &=& - \frac{3 \, \Omega_m \, H_0^2}{k^2} \int dz \, e^{-\tau(z)} \, g'(z) \, j_l[k \chi(z)]  \nonumber \\
W_l^{g_i}(k) &=& \int dz \,  b_{g_i}(k, z) \, \varphi_i (z) \,  D (z) \,  j_l[k \chi(z)]  \, ,
\ea
where  $\chi$ is the comoving distance and $j_l$ the spherical Bessel function.
These spectra are related to the auto- and cross-correlation functions (ACF and CCF), defined as
\ba
w^{g_i g_j} (\vartheta) &\equiv& \langle \delta_{g_i}(\hat {\mathbf {n}}) \, \delta_{g_j}(\hat {\mathbf {n}}') \rangle \nonumber \\
w^{T g_j} (\vartheta) &\equiv& \langle \delta_{g_i}(\hat {\mathbf {n}}) \, \Theta (\hat {\mathbf {n}}') \rangle \, ,
\ea
where the averages are carried over all pairs at a distance $\vartheta = |\hat {\mathbf {n}} - \hat {\mathbf {n}}'|$. They are derived from the spectra by Legendre transformation:
\be
w(\vartheta) = \frac{1}{4 \pi} \sum_{l=0}^{\infty}  \, (2 l + 1) \, C_l \, P_l [\cos (\vartheta)] \, ,
\ee
where $P_l$ are the Legendre polynomials.

\section {Data and Systematics}  \label{sec:data}
It is well known \cite{Ross13fnl,Huterer_calib,Pullen_qso} that the large-scale clustering of the LSS is easily contaminated by systematic distortions in the galaxy samples caused by observational issues, which may significantly bias the PNG results. Indeed, any spurious long-wavelength contaminant, produced in the Galaxy (extinction-causing dust, stars, synchrotron emission) or even in the atmosphere (seeing, airmass) will generally introduce extra large-scale power, which may na\"ively be attributed to PNG.
This issue only has a limited impact when studying the ISW correlation with the CMB, as  there is no particular reason why such  systematics should correlate with the primordial temperature fluctuations; however it has the potential to substantially bias any result using 
the large-scale angular auto-correlation of these galaxies.
However, as in the case of the galaxy-CMB correlations, cross-correlations between independent galaxy catalogs will be in general  less subject to such systematics, and can therefore provide estimates of the large-angle clustering that are more robust than the auto-correlations.

To minimize these effects, and to obtain robust measurements of PNG, we study a number of data sets previously considered in 
G12, including the main galaxy, LRG and quasar samples from the Sloan Digital Sky Survey (SDSS) \cite{York00}, as well as independent infra-red, radio and X-ray surveys.  We examine a number of these  in greater depth in order to address the impact of potential observational systematics on the large-scale clustering measurements.

\subsection{Data and General Considerations}

Below we discuss potential contaminants to our LRG, radio and quasar samples. 
The other data sets are unchanged from G12: The 2MASS infra-red catalog \cite{2006AJ....131.1163S} contains 415,459 galaxies after masking, has a median redshift $\bar z = 0.09 $ and a bias  $b = 1.3$ if it is assumed constant. The final imaging SDSS Data Release 8 (DR8) main galaxy sample \cite{Aihara:2011a} contains 30 million galaxies at $\bar z = 0.3$ and $b = 1.2$, while the HEAO survey of the X-ray background \cite{1987PhR...146..215B} has a broad redshift distribution of estimated $\bar z = 0.9$ and $b = 1.0$. See G12 for further details. 

We pixellate all maps using the \textsc{Healpix} \cite{2005ApJ...622..759G} scheme at low resolution $N_{\mathrm{side}} = 64$, corresponding to a pixel size of $\sim 50$ arcmin, since we are primarily interested in large scales. We measure the angular two-point correlation functions between two pixellated maps $a,b$ of weighted masks $f^a, f^b$ using the estimator
\be \label{eq:west}
\hat w^{ab} (\vartheta) = \frac{1}{N_{\vartheta}} \, \sum_{i,j = 1}^{N_{\mathrm{pix}}} \, \left( \frac{n^a_i}{\bar n^a} - 1  \right) \, \left( \frac{n^b_j}{\bar n^b} - 1  \right) \, f^a_i \, f^b_j \, ,
\ee
where $N_{\vartheta} = \sum_{i,j} f^a_i f^b_j$ is the weighted number of pixels at separation $\vartheta$,  $\bar n$ indicates the mean expectation value in each pixel and $N_{\mathrm{pix}}$ is the total number of pixels in a map. This generalizes Eq.~(10) of G12.

We account for the pixelization effects by smoothing the theoretical correlation functions with the \textsc{Healpix} window function in harmonic space. We checked that a higher resolution ($N_{\mathrm{side}} = 128$) leaves the results unchanged.

Note that the overlap between catalogs is significant, both in redshift and in footprint: this leads to significant cross-correlations and covariances in most cases, as also discussed in G08, G12.

\subsection {The BOSS CMASS Galaxies}

The MegaZ LRGs \cite{2007MNRAS.375...68C,2010arXiv1011.2448T,2011PhRvL.106x1301T} used in G08 and G12 have an excess of power on large scales with respect to the \LCDM~model \cite{2011PhRvL.106x1301T}. It was shown by Ref.~\cite{Ross:2011a} that the likely cause for this excess is a systematic contribution to the clustering from stellar contamination. 
Here, we have decided to instead use an updated catalog of LRGs at redshifts $z\sim0.55$, constructed from the final imaging DR8 of the SDSS \cite{Ross:2011a},  where the systematics are better understood.
We use the DR8 photometric redshift (photoz) sample rather than the data release nine (DR9) \cite{DR9} spectroscopic redshift sample recently used for multiple cosmological analyses \cite{BOSSDR9bao,Reid12bossDR9rsd,Ross13fnl,Samushia12boss,Sanchez12boss2pt,Scoccola12boss,Tojeiro12bossgrowth,Zhaobossneut}, because of the larger cosmological volume covered. Even though it relies on photometric rather than spectroscopic redshifts, it covers a usable area of $\sim$9,000 deg$^2$, rather than 3,300 deg$^2$ for the spectroscopic sample, giving an increase in volume that allows better large-scale clustering measurements.

\subsubsection{The Catalog and its Correlations}

The DR8 LRG sample was selected using the same color and magnitude cuts as the SDSS-III BOSS `CMASS' (constant mass) sample, defined in Ref.~\cite{2013AJ....145...10D}, using imaging data from the entire DR8 footprint. Here we will denote it as the CMASS sample. It used more than 100,000 BOSS spectroscopic redshifts for training to estimate photometric redshifts and evaluating the probability that an object is a galaxy.  (3\% of objects that are selected are stars.) A detailed analysis of possible systematic uncertainties was performed by Ref.~\cite{Ross:2011a}, including seeing effects, sky brightness and possible stellar contamination. The most serious systematic problem found with present photometric redshift catalogs from SDSS imaging was the loss of faint galaxies around stars (even relatively faint stars to $r\simeq20$). Ref.~\cite{Ross:2011a} corrected this issue, which clearly illustrates the need to be diligent about stellar contamination, particularly when cross-correlating with other data sets which may also include some contamination from galactic sources. 

\begin{figure}
\begin{center}
\includegraphics[width=\linewidth, angle=0]{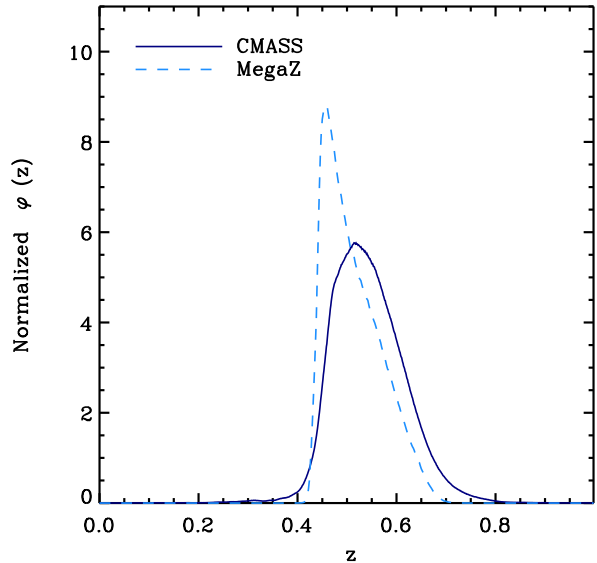}
\caption{Redshift distributions of the different LRG samples. The CMASS DR8 data are less peaked, but the difference in the distribution is small for the ISW purposes.}
\label{fig:LRGdndz}
\end{center}
\end{figure}

We use CMASS LRGs with $0.45 < z_{\mathrm{phot}} < 0.65$ and color term (see Eq.~13 in Ref.~\cite{Ross:2011a}), $c_{||} > 1.6$. This is an identical sample to that used in Refs.~\cite{Ho12,Seo12,dePutt12}, except that we do not further divide the sample by photometric redshift. Our map exploits the ``$A_{\mathrm{star}}$'' method (see Section 4.1 of Ref.~\cite{Ross:2011a}) to correct for the systematic effect of stars. We also correct for the offset between SDSS photometry in the North and South Galactic caps \cite{Schlafly11} (NGC/SGC from here on) using the method applied to obtain the ``$\Delta$South'' results of Ref.~\cite{Ross:2011a}.
The North/South offset for the CMASS LRGs was applied by Ref.~\cite{Ross:2011a} and it was recommended that it should be done for all analyses with the catalog. However, it mainly affects the measured clustering for $z_{\mathrm{phot}} > 0.6$, and it is a small effect; as it can be seen in Fig. 14 of Ref.~\cite{Ross:2011a}, this makes a small difference in the $0.6 < z_{\mathrm{phot}} < 0.65 $ bin only, so that it has a negligible effect in our case, where we do not cut in redshift bins.
We use the same mask for this sample as constructed in Ref.~\cite{Ho12}, but use a marginally more conservative Galactic extinction cut of $A_r < 0.18$ (compared to the fiducial $A_r < 0.20$), for consistency with our other data sets and previous analyses. This yields a sample with 801,226 LRGs occupying a footprint of 8,891 deg$^2$. The redshift distribution of the sample is centered around $\bar z \simeq 0.5$, as shown in Fig.~\ref{fig:LRGdndz}. By fitting the ACF, we have confirmed the bias $ b \simeq 2.1 $, found by earlier work~\cite{Ross:2011a}.

\begin{figure}
\begin{center}
\includegraphics[width=\linewidth, angle=0]{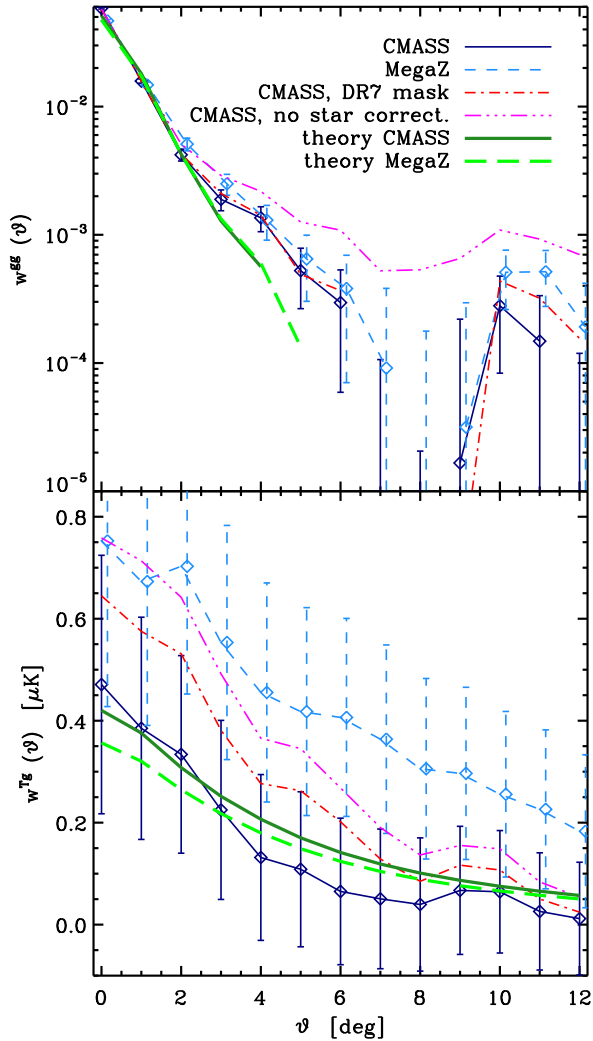}
\caption{Auto-correlation functions and cross-correlations with WMAP (top and bottom panels respectively) for the LRG data sets. The 1-$\sigma$ error bars are derived from Monte Carlos and are highly correlated; the data points have been slightly displaced in $\vartheta$ for clarity. Dark and light blue data show the CMASS and MegaZ LRGs, and the relative theoretical expectations from the WMAP7 best-fit cosmology (without PNG) and constant bias of 2.1 and 1.7 are drawn in dark and light green (ACFs negative at $\vartheta>5$ deg).  The red lines show the measured correlation functions for the CMASS data when the DR7 footprint mask is applied, while for the magenta line we have also removed the systematic correction for the effect of faint stars \cite{Ross:2011a}. A part of the differences between the LRG samples can be accounted for by these two effects.
} 
\label{fig:AshCCF}
\end{center}
\end{figure}

We calculate the auto- and cross-correlation functions of these LRGs and the WMAP7 internal linear combination (ILC) map using the estimator of Eq.~(\ref{eq:west}), binning the results in the range $0 \le \vartheta \le 12$ degrees. We also estimate the full covariance matrix using the Monte Carlo method MC2 of G08. We see the results in Fig.~\ref{fig:AshCCF}, where we also compare them with the correlations of the MegaZ data. While the ACFs are comparable at small scales, the excess large-scale power is significantly reduced in the CMASS data. Furthermore, the correlation of the CMASS data with the CMB is much lower than for the MegaZ, and 
on most scales comparable to the theoretical prediction for a WMAP7 best-fit \LCDM~cosmology (assuming a bias $b = 2.1$, and no PNG), remaining well within the (highly correlated) 1-$\sigma$ error bars. Any significant deviations could reveal either residual systematics, or PNG.

\subsubsection{Discussion}

The differences between the CMASS and MegaZ CCFs seen in Fig.~\ref{fig:AshCCF}  are large compared to their errors, and it is interesting to investigate the reasons for this discrepancy in more detail.

\paragraph*{\textbf{North-South Divide}}
The footprint of the CMASS data is larger than that of the MegaZ data (8,891 deg$^2$ compared to 7,400 deg$^2$ after masking), with most of the additional area resulting from imaging data in the SGC~\cite{Aihara:2011a}. As a first check, we tested whether the difference in the CCF signal is produced by the different sky coverage of the data. For this purpose, we measured the CCF of the CMASS data in the DR7 region only. We found that the signal increases significantly, as shown in Fig.~\ref{fig:AshCCF} (red line), although it still does not match the MegaZ result, which used DR7 (i.e., mostly the NGC data).

\begin{figure}
\begin{center}
\includegraphics[width=\linewidth, angle=0]{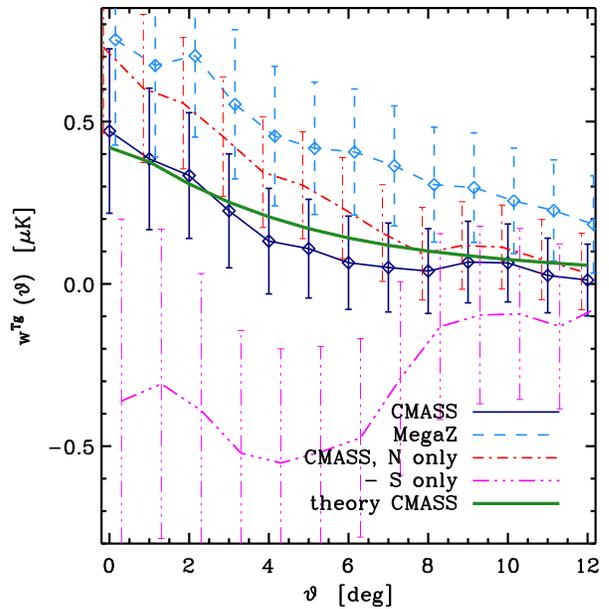}
\caption{North-South divide in the CMASS cross-correlation data. The results for the South galactic cap are contrary to the expected ISW signal, but the cosmic variance for this small area is large and it is potentially more subject to extinction systematics. 
}
\label{fig:NSdivide}
\end{center}
\end{figure}

We next examined whether the NGC and SGC footprints of DR8 give compatible ISW signals. As we can see in Fig.~\ref{fig:NSdivide}, this may not be the case, as the NGC of the DR8 data gives results which are in better agreement with the MegaZ catalog, but the SGC shows little or negative cross-correlation signal, and thus when added to the NGC data, lowers the overall CCF. 
However, given the relatively small coverage of DR8 in the SGC (1,300 deg$^2$), we should be wary of over-interpreting the results as the cosmic variance error on the ISW measurement is large. The difference between the NGC and the full DR8 data is more significant, but well within the expected statistical fluctuations given the size of the highly-correlated error bars in Fig.~\ref{fig:NSdivide}.

\begin{figure}
\begin{center}
\includegraphics[width=\linewidth, angle=0]{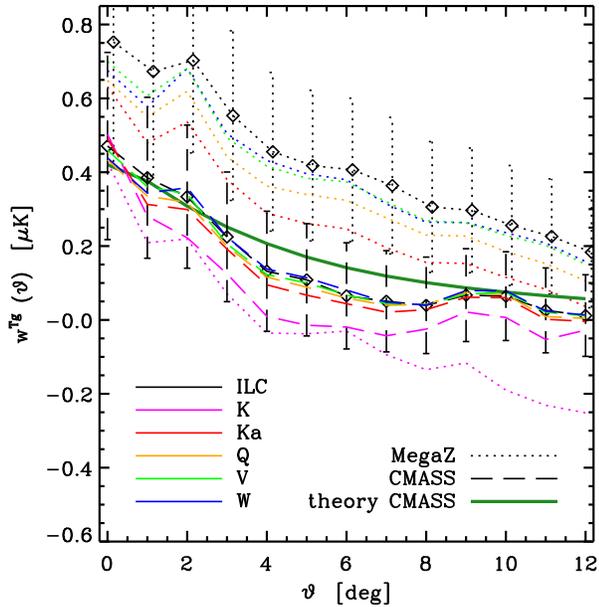}
\caption{Frequency dependence of the CCFs, for the MegaZ and CMASS data. We can see that the latter data are more stable across the whole frequency range.} 
\label{fig:freq8}
\end{center}
\end{figure}

\paragraph*{\textbf{Stellar Correction}}
We have also checked whether the new correction for the systematic effect of faint stars introduced by Ref.~\cite{Ross:2011a} can better explain the differences between MegaZ and CMASS. We can see from the magenta line in Fig.~\ref{fig:AshCCF} that indeed, if this correction is not applied to the CMASS data, the resulting CCF becomes even higher, close to the MegaZ result. However, even this effect is not enough to fully reproduce the earlier result on all angular scales.

\paragraph*{\textbf{Frequency Dependence}} 
An important property of the ISW signal is that it is expected to be independent of CMB frequency,  making any frequency dependence an indication of possible systematics. To quantify the level of contamination of the two LRG catalogs, we investigated in detail the frequency dependences of their CCFs, as shown in Fig.~\ref{fig:freq8}. Here we can see that the MegaZ result is more frequency dependent, while the CCF from the CMASS dataset is remarkably constant for all WMAP frequency bands, including the K and Ka WMAP bands which are most affected by residual galactic emission. This robustness, unmatched in any of the other galaxy catalogs from G12, is fully consistent with the ISW interpretation of the observed cross-correlation signal and appears to confirm that the stellar contamination in the CMASS data is negligible.  From this test, we conclude that the CMASS catalog is the most robust data set available for our purposes.

\paragraph*{\textbf{Extinction} }
As discussed in G08 and G12, dust extinction can be a major source of systematic uncertainty in ISW measurements. This is due to reddening from our Galaxy altering the inferred large-scale distribution of galaxies, which then introduces spurious correlations with the CMB, which may also have residual galactic emission. For this reason, we have carefully checked the effect of extinction on the measured CCF.

\begin{figure}
\begin{center}
\includegraphics[width=\linewidth, angle=0]{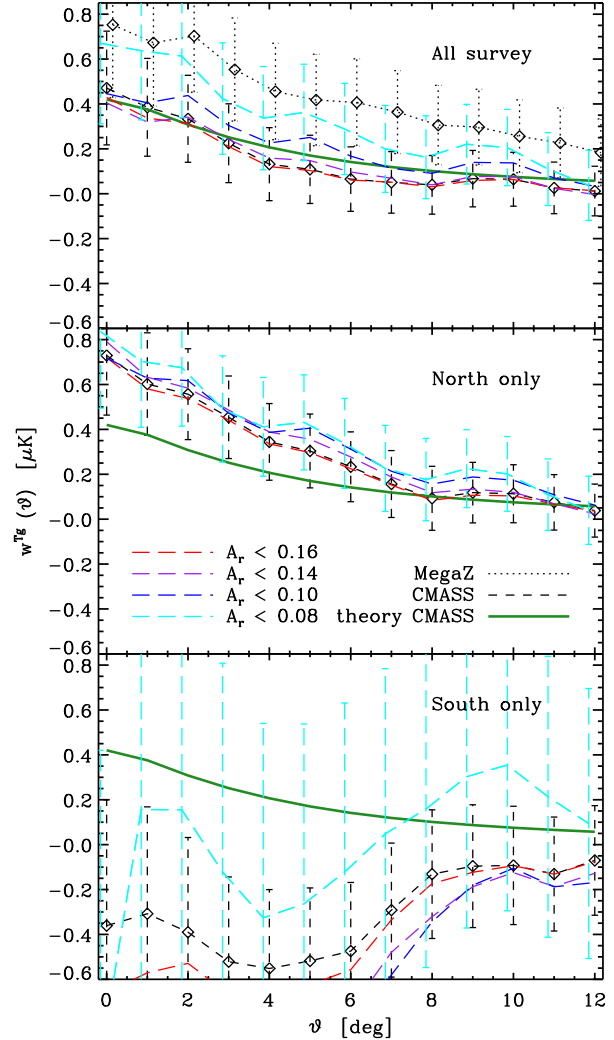}
\caption{Extinction dependence of the CCFs for the CMASS LRGs. The top panel shows the result for the whole surveyed area, while the middle and bottom panels present results from the NGC and SGC respectively. The black lines refers to our baseline cut at $A_r < 0.18$. The greater variation in the SGC reflects both its smaller area and the higher average reddening in this region, and it is consistent within the errors.
}
\label{fig:redd}
\end{center}
\end{figure}

Extinction can be corrected by excluding the most affected regions from the analysis. In  Ref.~\cite{Ross:2011a}, areas with a reddening $A_r > 0.20$ magnitudes were excluded, while the cut in G12 was marginally stricter ($A_r > 0.18$) and we have applied the latter in the present analysis. Fig.~\ref{fig:redd} shows how the CCFs vary when lowering the amount of reddening which is tolerated. When this cut is made stricter, the measured signal increases both for the full DR8 and NGC-only areas.
The average extinction is higher in the SGC; the average in the whole footprint, calculated before applying the extinction cut itself, is $\bar A_r = 0.12 $; this reduces to $\bar A_r = 0.10 $ in the NGC and increases to $\bar A_r = 0.18 $ in the SGC.
The sky coverage decreases significantly for higher extinction cuts, going from $f_{\mathrm{sky}} = 0.22$ for  $A_r > 0.18$ to $f_{\mathrm{sky}} = 0.11$ for  $A_r > 0.08$, of which only $f_{\mathrm{sky}} = 0.01$ remains in the South. This means that the error bars, which scale as $\sqrt{1/f_{\mathrm{sky}}}$ due to sample variance, are significantly higher. As we can see in Fig.~\ref{fig:redd}, this increase in the errors makes the results for the different extinction cuts statistically consistent.

\subsubsection{Conclusion}

From the analyses performed in this section, we conclude that the CMASS data set is the most robust available for large-scale structure and cross-correlation measurements, as it appears to have low stellar contamination \cite{Ross:2011a}. We also observe that the lower CCF from the CMASS data is partly due to a lower CCF in the SGC portion of the DR8 data; while consistent with cosmic variance, this could also be potentially connected to higher extinction effects in the SGC region.  

\subsection{NVSS Systematics} \label{sec:NVSS}

The NRAO VLA Sky Survey (NVSS) \cite{1998AJ....115.1693C} produced a catalog of radio-galaxies covering the full sky at declinations $ \delta > -40 $ deg. The clustering properties of these sources have been measured by several authors \cite{2002MNRAS.329L..37B,2002MNRAS.337..993B,2004MNRAS.351..923B, 2006MNRAS.368..935N}, who found excess power on the largest scales compared with the \LCDM~predictions and the analyses of other data sets.  
Its combination of large sky coverage and redshift depth makes the NVSS well-suited for the measurement of ISW cross-correlations and  this has been one of the most studied data sets in this context \cite{2002PhRvL..88b1302B,2004Natur.427...45B,2004ApJ...608...10N,2006PhRvD..74d3524P,2008MNRAS.384.1289M,2008MNRAS.386.2161R,2010A&A...520A.101H,2012arXiv1203.3277S}. Furthermore, these data have been used for other studies, such as to detect CMB lensing \cite{2007PhRvD..76d3510S} and to constrain inhomogeneous models \cite{2010MNRAS.403....2S}.  

Unfortunately this catalog is affected by some well-known issues. There is a large uncertainty in the redshift distribution of the sources, which can make its cosmological interpretation ambiguous.  For many applications, this is not a serious issue; for example, it was shown in G12 that different assumptions for the distributions do not significantly affect the ISW measurements, given the size of the expected errors. 

\begin{figure}
\begin{center}
\includegraphics[width=\linewidth, angle=0]{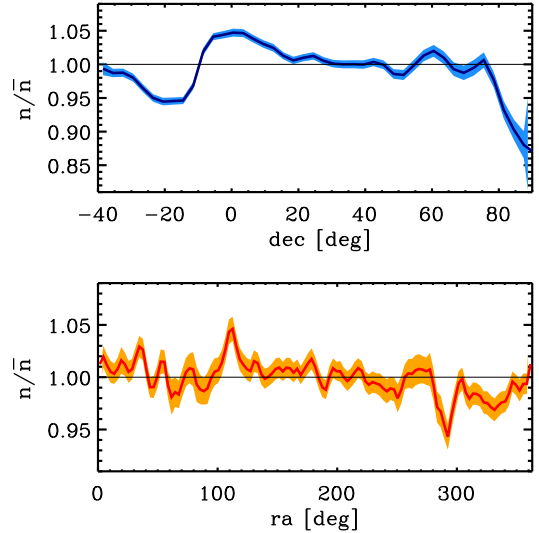}
\caption{Systematic density fluctuations in NVSS. The upper panel shows the well-known variation as a function of declination, while the lower panel presents a weaker deviation we have found as a function of right ascension. The data have been smoothed with a Gaussian beam of r.m.s. $\sigma = 5 $ deg. The colored bands show the Poisson $1-\sigma$ regions, which are largest near the poles.}
\label{fig:nvss_radec}
\end{center}
\end{figure}

More worryingly, the sample contains clear variations in the density of sources with declination that are significantly greater than those expected for a statistically isotropic universe \cite{2002MNRAS.329L..37B,2002PhRvL..88b1302B}.  In particular, fewer sources are detected at $\delta < -15 $ deg, as can be seen in the upper panel of Fig.~\ref{fig:nvss_radec}. In addition, we have found a similar albeit smaller issue in right ascension: the number density has a roughly linear decrease, reaching a minimum at r.a. $> 250 $ deg; this can be seen in the lower panel of Fig.~\ref{fig:nvss_radec}. The data points at r.a. $ = 0/360$ match as the distinction is purely artificial.

The main effect of such systematics, and of any long-wavelength noise in the data, will be to add spurious signal on the large-scale power spectrum (or ACF). Indeed the presence of spurious large-scale clustering power above an assumed cosmological model is often taken to be a clear indicator of uncorrected systematics in the data; however, we must be careful not to do so here as we are trying to measure, or limit, a large-scale PNG signal from the data.

In previous analyses, a number of methods have been used to correct the declination systematics: in some real-space analyses \citep{2002PhRvL..88b1302B,2004Natur.427...45B,2010MNRAS.403....2S}, including G08 and G12, the data were subdivided in declination bands, and expected densities were rescaled to match each. Other authors \citep{2002MNRAS.329L..37B,2004MNRAS.351..923B, Xia:2010b} found that this problem was minimized by imposing a strict flux cut of $F > 10 $ mJy; this however discards a large fraction of sources. Finally, others \citep{2007PhRvD..76d3510S,2013MNRAS.tmp.2142L} working in harmonic space addressed the issue by assigning infinite variance to the $m=0$ modes of the data, effectively discarding any azimuthally-invariant mode.

The method of fitting sub-bands in declination effectively assumes that any declination offsets are smooth and fit by the particular distribution of bands chosen. Going beyond the earlier analyses in G08 and G12, we have found that the signal fluctuates depending on the size chosen for these bands, leading to an uncertain auto-correlation. 
This suggests that we need a more robust technique for correcting for these effects if we are to use this sample to robustly measure PNG.
However, the cross-correlation with WMAP is fairly stable to such corrections: we show below in Section~\ref{sec:resultsLCDM} that the G12 results on the significance of ISW detection do not change significantly.

We have compared these existing methods for mitigating the NVSS systematics, and have also tested a new method, modelling the mean source density with an arbitrary mask that is assumed to be a separable function of the r.a. and dec coordinates.  Any systematics or true fluctuations in the associated modes are absorbed into the inferred mask, so the inferred power is suppressed. We tested these methods with mock data to which we applied the observed r.a. and dec systematics, and found that this method does indeed allow us to remove their effects, and in particular the resulting large-angle power that is produced in the ACF. However, such corrections are problematic when constraining PNG, as they suppress the true power as well; when we applied these methods to non-Gaussian mocks, we found that the large-angle correlations produced by $\fnl$ are also removed to a significant degree by this systematic mitigation procedure in ways that are difficult to model in real space. Since removing this type of systematic in real space without also removing any primordial signal is difficult, we decided to adopt the most conservative approach and not use the NVSS ACF to constrain cosmology nor PNG. The cross-correlations with other density tracers and the CMB are instead expected to be reliable, as long as these other data sets are not affected by correlated systematics, and we therefore retain these measurements in our analysis.

If we include r.a. and dec systematic fluctuations in our mocks at a level consistent with the NVSS observations, then the clustering seen in the mocks is a good match to that in the observational data, as shown below in Fig.~\ref{fig:gg}. We use these contaminated mocks with the density fluctuations to estimate the covariance matrix, thus estimating the data errors using a set of mock ACFs with amplitude in agreement with the actual raw NVSS ACF.

\begin{figure}
\begin{center}
\includegraphics[width=\linewidth, angle=0]{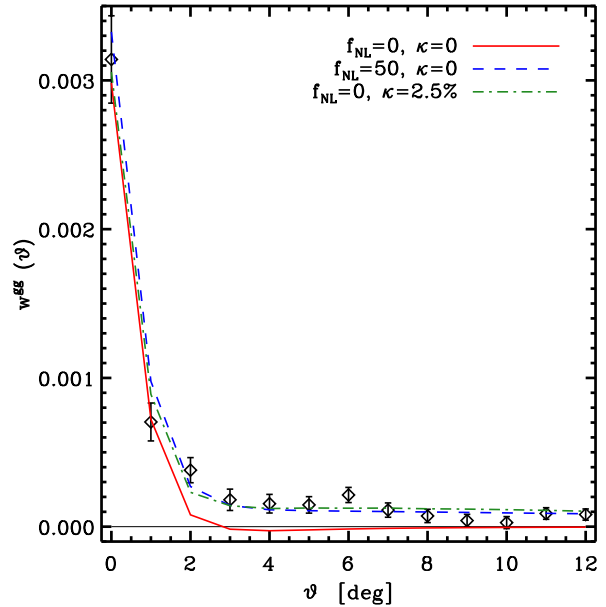}
\caption{Large-scale clustering of the quasars and stellar contamination. We plot in red (solid) the prediction for a model with no stellar contamination and no PNG. The green (dot-dashed) and blue (dashed) lines correspond to models with stellar contamination and with PNG, which are both able to explain the observed excess power (black data points). Constant linear bias is assumed.} 
\label{fig:qso_acf}
\end{center}
\end{figure}

\begin{figure}
\begin{center}
\includegraphics[width=\linewidth, angle=0]{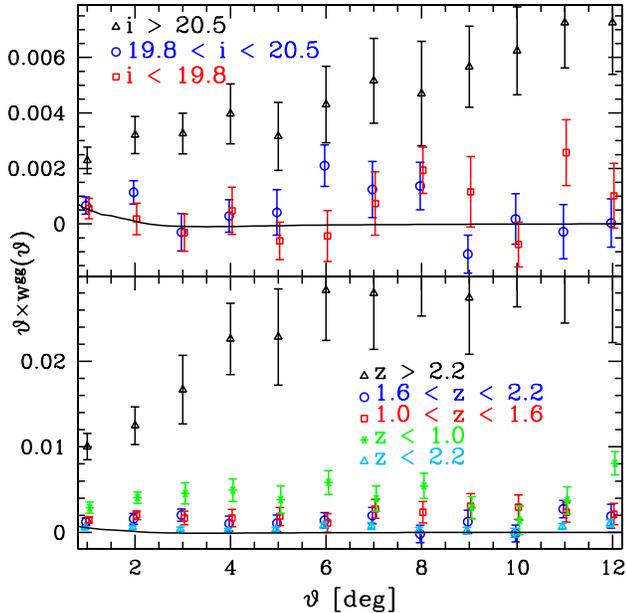}
\caption{\emph{Top panel:} The angular auto-correlation function, ACF, multiplied by the angular scale $\vartheta$, when the DR6 quasar sample is sub-divided based on $i$-band magnitudes. The solid black line is the prediction for the standard \LCDM~model. \emph{Bottom panel:} The measured ACF when the DR6 quasar sample is split based on photometric redshift, $z$.}
\label{fig:qsomagz}
\end{center}
\end{figure}

\subsection{Quasar Systematics} \label{sec:QSO}

Quasar samples are promising for future PNG measurements, as they are easily observed to high redshifts and are highly biased tracers of the matter distribution. The DR6 photometric quasar catalog has been used previously to constrain PNG, and Ref.~\cite{Xia:2011a} found that these data alone yield $\fnl = 62 \pm 26$ (1$\sigma$). However, this result is in conflict with Ref.~\cite{Slosar:2008a}, who found $\fnl = 8^{+26}_{-37}$ (for their fiducial `QSO' case): they used an earlier sub-set of the DR6 data from the SDSS DR3 \cite{DR3}. Ref.~\cite{Pullen_qso} (and more recently Ref.~\cite{2013MNRAS.tmp.2142L}) thoroughly tested the DR6 quasar sample against potential contaminants, and found significant effects associated with stellar density, the stellar color locus, airmass, and seeing. They concluded that the sample was not fit to use for $\fnl$ measurements.  Here we further investigate this issue, and argue that the quasar data should only be included through cross-correlations with other surveys. 
Future spectroscopic quasar samples will be immune to many of these issues. 

\begin{figure*}
\begin{center}
\includegraphics[width=\linewidth, angle=0]{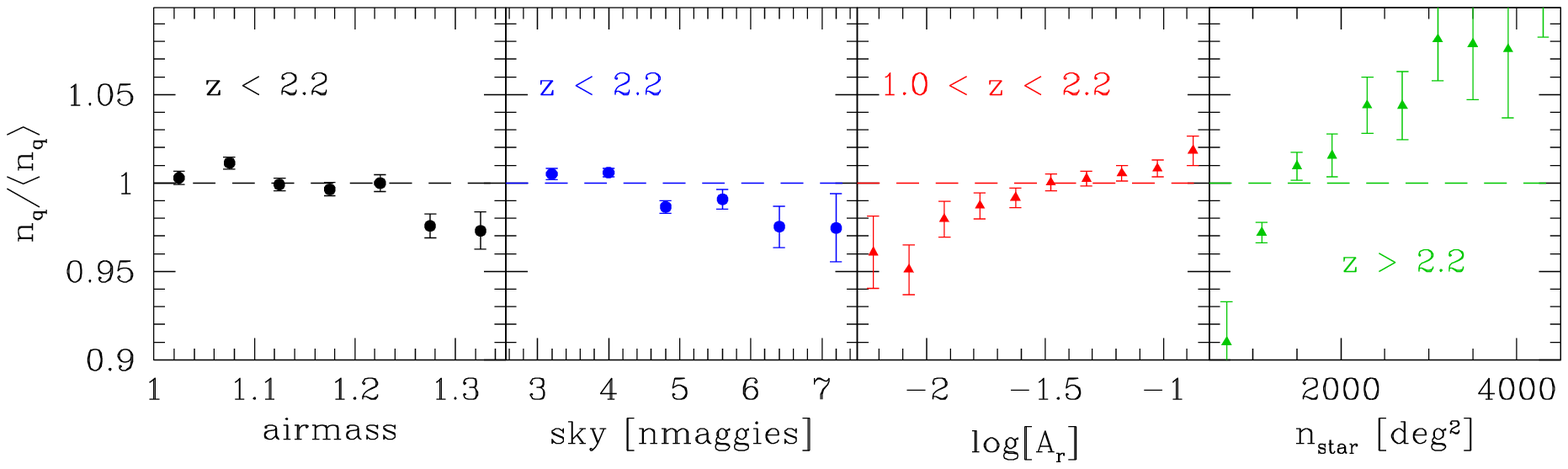}
\caption{Observed relationship between the number density of DR6 quasars $n_q$ with respect to the average $\langle n_q \rangle$ 
and a choice of systematics: airmass, sky brightness, galactic extinction in the $r$-band, and stellar density, in different redshift ranges.}
\label{fig:qso_sys}
\end{center}
\end{figure*}

The fiducial photometric quasar sample we use is drawn from the SDSS data release six (DR6) \cite{DR6}. The quasar catalog \cite{RichardsDR6} contains more than one million objects selected via a non-parametric Bayes classifier -- kernel density estimate (NBC-KDE) method, including both UV-excess (UVX) and high-redshift samples, up to a magnitude limit of $i = 21.3$. The total footprint is 8417 deg$^2$. Nevertheless, to reduce contamination as much as possible, we restrict the analysis to the UVX sample, as in G08 and G12, and we reduce the footprint with a stricter extinction cut, discarding areas with $A_r > 0.14$, so that we actually use 502,565 sources covering 6,912 deg$^{2}$.

Photometric quasars are particularly prone to stellar contamination, as their faint and compact nature matches those of stars. If we assume a stellar contamination fraction given by  $\kappa$, the expected auto-correlation function will be given by 
\be \label{eq:kappa}
w^{\mathrm{obs}}(\vartheta) = (1 - \kappa)^2 \, w^{\mathrm{qso}}(\vartheta) + \kappa^2 \, w^{\mathrm{star}}(\vartheta) \, ,
\ee
where we estimate the stellar ACF $w^{\mathrm{star}}(\vartheta) $ by measuring the clustering of stars from the SDSS survey. The distribution of stars over the DR6 footprint is distinctly anisotropic since they trace the structure of the Galaxy, and the auto-correlation of stars is therefore significant at large scales.
A residual contamination of the order of $\sim 2.5 \%$ could explain the observed plateau in the ACF as shown in Fig.~\ref{fig:qso_acf}, as could $\fnl \sim 50$. We found from both the local relationship between quasar density and stellar density and also by directly measuring the cross-correlation function between the quasars and a catalog of SDSS stars that the actual contamination fraction is only $\sim 1\%$ in these data. The quasar ACF data thus suggest there are either further spurious fluctuations in the quasar density field, or $\fnl$ is non-zero (as suggested by Ref.~\cite{Xia:2011a}).

It is reasonable to assume that sub-samples at fainter magnitude should be less reliable and more severely affected by systematics. We therefore measured the ACF with cuts on $i$-band magnitude and we display the results in the top panel of Fig.~\ref{fig:qsomagz}.  Indeed, the ACF of the faintest data ($i > 20.5$; black triangles) displays much larger clustering amplitudes than the brighter sub-samples (and the difference is inconsistent with the change in linear bias and a non-zero $\fnl$). However, we are unable to isolate a particular magnitude at which we find that the systematic relationships become negligible and we do not find that the $\fnl$ constraints from the DR6 quasar ACF become stable for samples brighter than any given $i$-band magnitude.

The bottom panel of Fig.~\ref{fig:qsomagz} displays the ACF when we split the quasar sample by photometric redshift. Both the $z > 2.2$ (black triangles) and $z < 1.0$ (green stars) samples display significant large-scale clustering, that we can attribute to, in large part, significant stellar contamination in these subsamples (confirmed both via cross-correlation and local relationships with stellar density). However, we find that cutting by photometric redshift can introduce systematic fluctuations, as would be the case if there were any relationship between the estimated redshift and a potential systematic. This is illustrated by the fact that the ACF of both the $1.6 < z < 2.2$ (blue circles) and $1.0 < z < 1.6$ (red squares) samples has larger amplitudes than the $z < 2.2$ (cyan triangles) ACF, despite the fact that the $z < 2.2$ sample contains quasars with $z < 1.0$ and thus has significant stellar contamination. Thus, while the $1.0 < z < 2.2$ sample removes data with stellar contamination, the cuts on redshift appear to induce other systematic fluctuations. Some examples are shown in Fig.~\ref{fig:qso_sys}: e.g. we find a significant relationship with Galactic extinction $A_r$, when we cut the sample to $1.0 < z < 2.2$, but find no such relationship when no redshift cut is applied. We further find significant correlations with air mass, sky brightness, and especially at high redshift, with the surface density of stars.

\begin{figure*}
\begin{center}
\includegraphics[width=0.76\linewidth, angle=0]{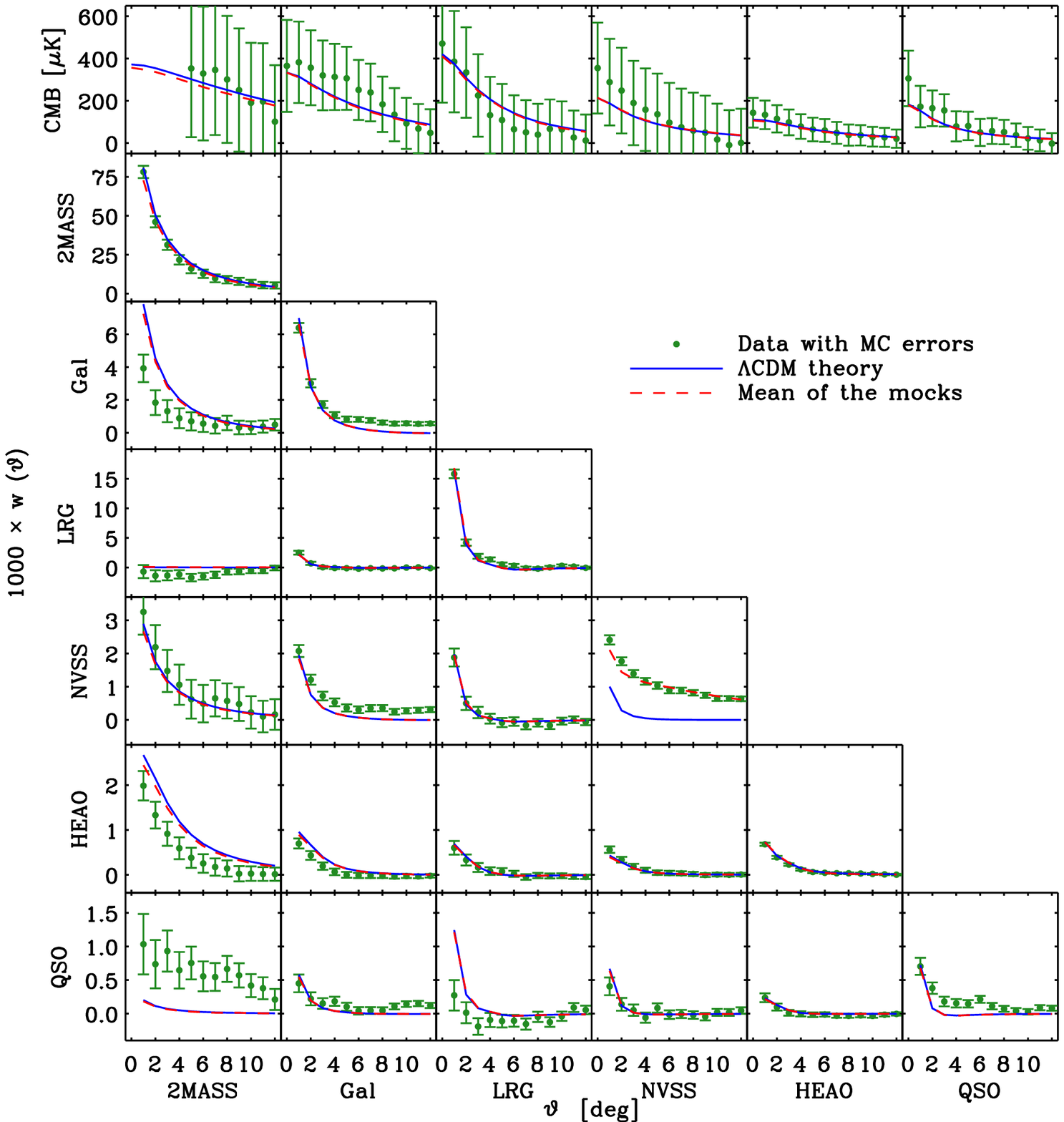}
\caption{Complete set of the two-point functions we use. The top row shows the CMB-galaxy correlation functions, while the remaining panels are the galaxy-galaxy correlations. Error bars are from 10,000 Monte Carlos, whose means are the red dashed lines, and the blue line is the standard \LCDM~cosmology from WMAP7, with constant biases (not a fit to these data). The ACF of the raw NVSS data presents a significant excess power with respect to the \LCDM~expectations, which is modelled by adding to the mocks the r.a. and dec density fluctuations observed in the data.}
\label{fig:gg}
\end{center}
\end{figure*}

We have not studied the DR6 quasar sample as thoroughly as Ref.~\cite{Pullen_qso}, but we have, independently, reached a similar conclusion: the DR6 quasar ACF should not be used to obtain $\fnl$ constraints. We have found that significant correlations exist with potential systematics, such as stellar density, Galactic extinction, and airmass, exist in excess of those expected. Also, these relationships depend non-trivially on the redshift and magnitude of the quasar sample that is selected. The effect of $\fnl$ on the quasar density field will also depend on redshift and the bias of the sample, and we are therefore unable to address systematic concerns using the methods outlined in Refs. \cite{Ross:2011a,Ho12}.

However, we do not expect these issues to be correlated with other samples, and should be able to trust correlations between the quasars and other data sets. In particular, the quasars have a large overlap in redshift with the NVSS data. Potential SDSS systematics, such as airmass and seeing, are survey-specific and should thus have no correlation with NVSS data. In addition, we find no correlation with NVSS data and potential systematics (Galactic extinction, stellar density, synchrotron emission) that trace the structure of the Galaxy. Further, we trust correlations between the quasars and the LRGs, as the LRG sample has already proven to be robust to systematic fluctuations. Thus, while we do not consider the quasar ACF as a reliable probe of PNG, we will exploit the external correlations between the quasars and the other data sets.
Also in this case, this includes the cross-correlation with the CMB, which for the same reasons should be relatively free from contamination, as also confirmed by its fequency independence shown in G12.

\begin {table*}
\begin {center}
\begin{tabular}{| c | c | c | c | c |}
\hline
 & Parameter                             &  Description        &  Prior range & In final runs \\
\hline
Cosmology &$ \omega_b  \equiv \Omega_b h^2 $     & baryon energy density      &  $[0.015,0.035]$  & Yes \\
~ &$ \omega_c  \equiv \Omega_c h^2 $     & dark matter energy density &  $[0.05,0.2]$  & Yes \\
~ &$ \theta $     & sound horizon at the last-scattering surface &  $[0.8,1.2]$   & Yes\\
~ &$ \tau $     & optical depth &  $[0.01,0.25]$ & Yes \\
~ &log $(10^{10} A_s )$     & amplitude of primordial perturbations at $k_{\mathrm{piv}} = 0.002 h/$Mpc &  $[2.9,3.5]$   & Yes\\
\hline
Our focus ~ &$ \fnl $     & amplitude of skewness-type primordial non-Gaussianity &  $[-200,200]$    & Yes \\
~ &$ \gnl $     & amplitude of kurtosis-type primordial non-Gaussianity &  $[-5 \cdot 10^6,5 \cdot 10^6] $  & No \\
~ &$ \anl $     & scale-dependence of bias &  $[0,4]$   & No \\
~ &$ w $     & dark energy equation of state &  $[-2.5,-1/3]$    & No \\
\hline
Nuisance &$ A_{\mathrm{SZ}} $     & amplitude of the SZ CMB template &  $[0,2]$  & Yes \\
~ &$ b_0^i $     & 6 bias nuisance parameters for each catalog &  $[0,3]$  & Yes \\
~ &$ \gamma_i $     & slope of bias evolution for each catalog &  $[0,3]$  & No \\
~ &$ \kappa_i $     & 3 stellar contamination nuisance parameters for each SDSS catalog &  $[0,0.1]$   & Yes \\
~ &$ \beta_{ij} $     & 15 $dN/dz$ nuisance parameters for each pair of catalogs &  Gaussian priors   & No \\
~ &$ \alpha_{\mathrm{HEAO}} $     & 1 PSF nuisance parameters for the HEAO catalog &  $[0.0001,0.001]$  & Yes \\
\hline
Derived & $ \Omega_{\Lambda} $     & dark energy energy density &  ---  & --- \\
~ &$ T_0 $     & age of the Universe &  ---  & --- \\
~ & $ \Omega_{m} $     & total matter energy density &  ---  & ---\\
~ & $ \sigma_{8} $     & LSS amplitude of density fluctuations &  ---  & ---\\
~ & $ z_{\mathrm{re}} $     & redshift of reionization &  ---  & ---\\
~ & $ H_0 $     & Hubble constant &  ---  & ---\\
\hline
\end{tabular}
\caption{Summary of the parameters used in the likelihood runs, and their prior ranges. We checked that our posterior results are never affected by the edges of the priors. We specify in the last column the parameters that we are using for our final $\fnl$ runs (`conservative' and `fair').}
\label{tab:params}
\end{center}
\end {table*}

\section {Modeling the Data} \label{sec:method} 

\subsection{Data Considered}

We have discussed six different large-scale structure data sets, which yield six auto-correlations, fifteen cross-correlations and six correlations with the WMAP CMB temperature.  Our final data set is shown in Fig.~\ref{fig:gg}, including
the galaxy-CMB cross-correlations and the galaxy-galaxy correlations between all the catalogs. As any of the corrections to the NVSS r.a. and dec fluctuations that we have tested can potentially remove the PNG signal, we are conservatively not using the NVSS ACF. We use the raw NVSS data without any correction, to measure external cross-correlations as described below, since the other data sets should not correlate with these systematic fluctuations.
While random correlations between the NVSS systematics and the other data may in principle occur by chance, we have checked that our results remain fully consistent and are not biased if instead we apply a systematic correction procedure to NVSS.

In all cases below, we discard the first five angular bins of the 2MASS-CMB CCF, which is believed to be contaminated by the SZ effect. We also discard the zero-lag data points in the density-density correlations based on contamination from non-linear structure formation, which we are not modeling, and from possibly inaccurate shot-noise subtraction in the correlations between different catalogs, due to the unknown fraction of common sources between the catalogs.

Given the potential impact of systematic contaminants, it is useful to define the following three sub-sets of data (we exclude in all cases the 0-lag bins of the density-density correlations and the first five data points in the 2MASS ISW measurement as described above):
\begin{itemize}
\item The \textbf{`na\"{i}ve'} data set is the full combination. As we are conservatively using the raw NVSS data, without any correction for the r.a. and dec systematics, we can not use the NVSS ACF for any cosmological constraints. So this data set now includes 26 correlation functions.
\item The \textbf{`fair'} sample contains {25 correlations, excluding the auto-correlations of NVSS and QSO, which are expected to be overall the least reliable parts of our data set, due to the high risk or residual systematics (as discussed above).}
\item The \textbf{`conservative'} set includes only what we view as the most reliable correlations. We exclude all the auto-correlations except for the LRG case, as auto-correlations are expected to be more prone to systematics than the cross-correlations. Further, we exclude any data from 2MASS and from the main SDSS galaxies; as they are not expected to contribute significantly to the PNG measurements due to their low bias, we feel it is better to exclude them rather than risk any systematic contamination that they might bring. Finally, we also exclude the LRG-quasar correlation: as both samples are derived from SDSS, there is a limited risk of residual correlated systematics, particularly related to stellar contamination. The resulting set is consists of ten correlations only.
\end{itemize}
We will discuss the results from these samples
in the following section.
Finally,  it is worth noting that we discard samples by effectively setting the covariance for these data to a high value. The likelihood estimation from the inverse of the covariance matrix will then simply ignore the discarded points.

In addition to our data, we will consider also the CMB data from WMAP7 \cite{Larson:2011a} and in some cases the luminosity distances to the Union2 compilation of Type Ia Supernovae \cite{2010ApJ...716..712A}.

We calculate the theoretical correlation functions using a modified version of \textsc{Camb} \cite{Lewis:2000a} (validated with a fully independent code for accuracy). The covariance matrix is generated using the Monte Carlo method described in G12, now extended to the whole data set (we have $n = 351$ data points, so this is the dimension of the full covariance).
We base our Monte Carlo realizations on a fiducial flat \LCDM~model without PNG, which is appropriate since we do not find posterior evidence for PNG; if such evidence were found, we would need to re-calculate our likelihoods with a new covariance matrix, based on a non-Gaussian model. To calculate the covariance, we use the linear bias parameters fitted from the auto-correlation functions and assumed constant in redshift. In the case of NVSS, we fitted the bias to the ACF when the systematics are subtraced, which is the only step for which we use the corrected NVSS data; the obtained bias value agrees with the external correlations of NVSS, and furthermore the fact that the mock samples reproduce the measured clustering also for the ACF implies the bias we used is reasonable.

\subsection{Gaussian Bias and Other Nuisance Parameters}
In addition to the standard cosmological parameters, we must include other nuisance parameters to account for the systematics and other uncertainties, over which we marginalize.  These describe our assumptions about the intrinsic bias (in the absence of PNG effects), 
uncertainties in the redshift distributions of the surveys (and particularly their overlaps) and other effects. 

For most catalogs $i$, we assume an intrinsic (Gaussian) bias evolution of the form:
\be
b_1^i(z) = 1 + \frac {b_0^i - 1}{D^{\gamma_i}(z)} \, ;
\ee
we assume  $\gamma_i = 2$ in most cases \cite{1998MNRAS.299...95M}; this bias approaches unity in the future, and is therefore appropriate for flux-limited samples. However, for the quasars we instead assume
\be
b_1^{\mathrm{QSO}}(z) = \frac {b_0^{\mathrm{QSO}}} {D^{\gamma_{\mathrm{QSO}}}(z)} \, ,
\ee
which instead approaches zero in the future. This is consistent with observations of quasars \cite{Pad09qso, White12qso}, that suggest that the preferred halo mass of quasars is $\sim10^{12}M_{\odot}$, independent of redshift. Setting $\gamma_{\mathrm{QSO}} = 1.6$ approximately satisfies this constant-mass condition for $0.43 < z < 2.4$, which spans the median redshifts of Refs.~\cite{Pad09qso,White12qso}, and is the span of redshifts at which we may reasonably expect to obtain meaningful information from the DR6 quasar sample.
We are relatively uncertain of the bias evolution of the NVSS sample, which has a broad redshift distribution but also the potential to provide strong PNG constraints due to its large sky coverage. We therefore have tested allowing a free slope $\gamma_{\mathrm{NVSS}}$ for this sample and found that this results in negligible changes in our constraints. This suggests that our results are not strongly affected by uncertainties in the bias evolution relationship.

In addition we allow the possibility to add 15 extra parameters $\beta_{ij}$ to account for uncertainties in the redshift distributions $\varphi_i(z)$.  A factor arises for each pair of catalogs, and they are introduced in the density-density cross-correlations so that:
\be
w^{g_i g_j}_{\mathrm{obs}} (\vartheta) = \beta_{ij} \, w^{g_i g_j} (\vartheta) \, .
\ee
Effectively, the uncertainty in $\varphi_i(z)$ becomes more important when the overlap between two surveys is smaller. For this reason, we introduce a Gaussian prior, with mean $\mu_{\beta} = \langle \beta_{ij}\rangle = 1$ and a variance $\sigma^2_{\beta}$ which depends on the fiducial survey overlap, such that the r.m.s. is $\sigma_{\beta} = 0 $ when the overlap is total (the ACFs), and $\sigma_{\beta} = 2 $ when the overlap is expected to be small. We linearly interpolate the prior between these two values. We also checked that the results on PNG do not change significantly when doubling the r.m.s. of these priors.

We also add a free stellar contamination fraction $\kappa_i$ for the three samples derived from SDSS, used as described for the quasars in Eq.~(\ref{eq:kappa}). 
We finally add one extra parameter for the HEAO catalog: an amplitude describing the clustering-independent auto-correlation due to source shot noise spread over the large PSF of the instrument $\alpha_{\mathrm{HEAO}}$ \cite{2002ApJ...580..672B}.

\begin{figure}
\begin{center}
\includegraphics[width=\linewidth, angle=0]{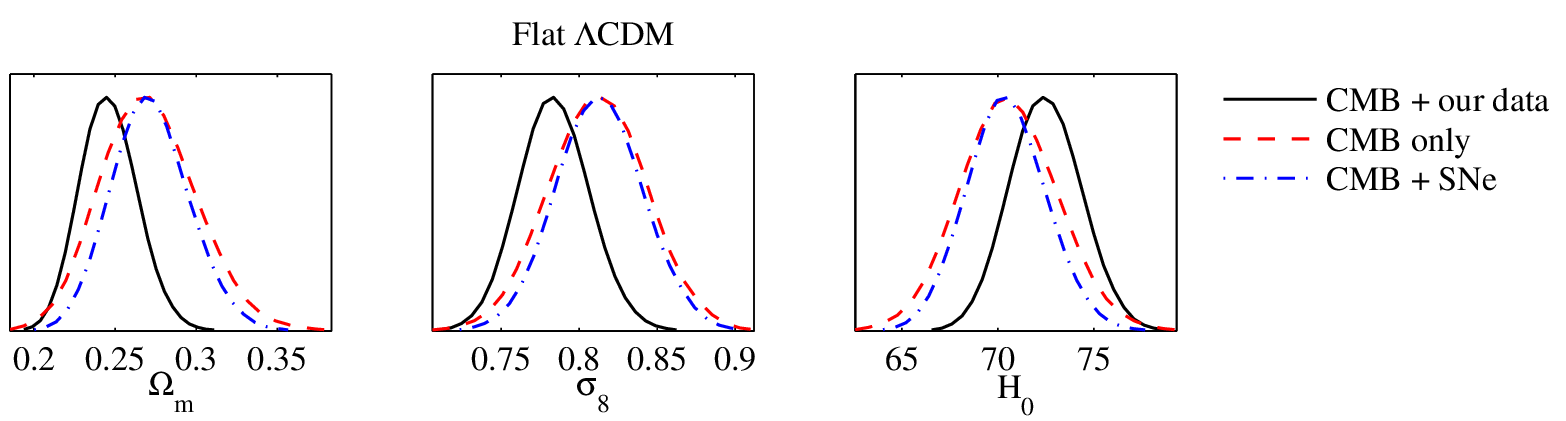}
\includegraphics[width=\linewidth, angle=0]{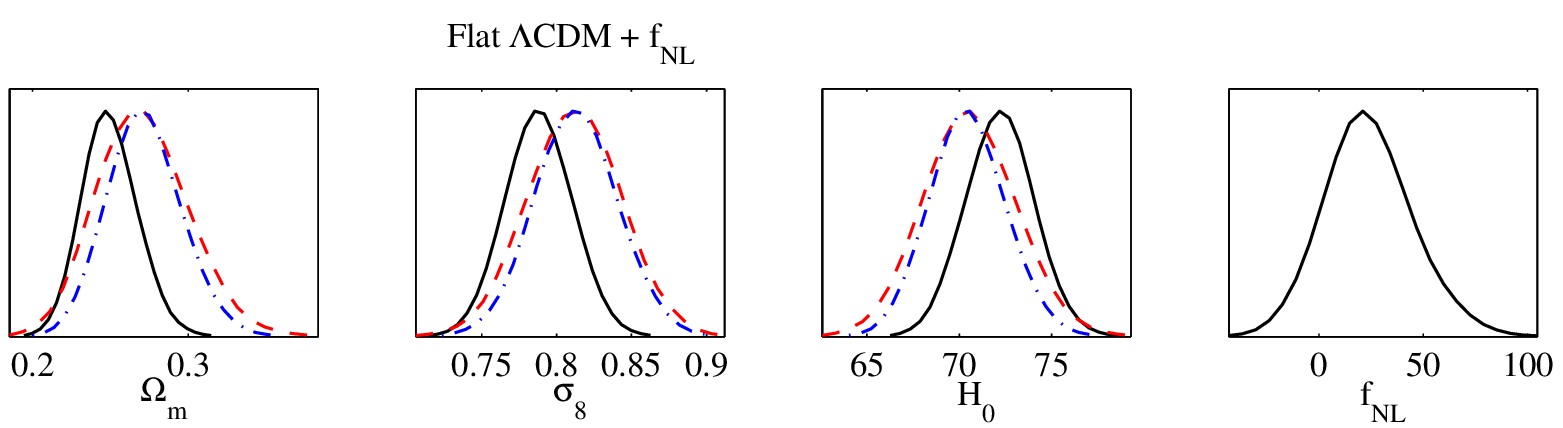}
\includegraphics[width=\linewidth, angle=0]{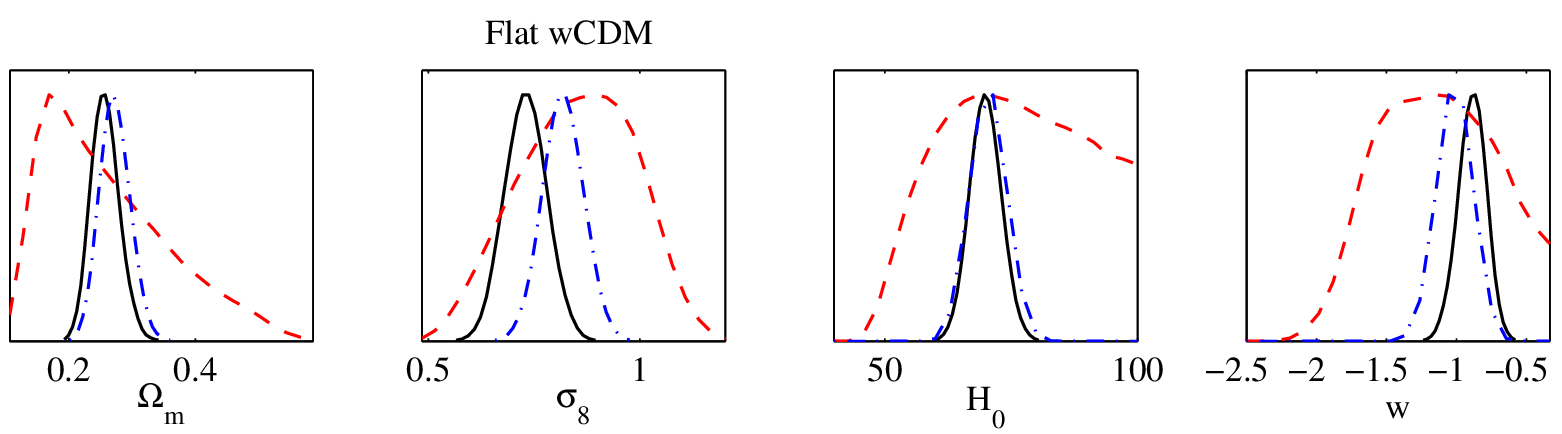}
\caption{Marginalized posterior probability distributions on a choice of cosmological parameters for different models. \emph{First row:} The flat \LCDM~case. The results are in agreement with the concordance model, but we can see that our data (the `fair' data set is used) shift the peak of the distribution towards models with more dark energy.  \emph{Second row:} Same, with the addition of local $\fnl$. (The CMB temperature power spectrum can not constrain PNG and is ignored.) The standard cosmological parameters do not change, and $\fnl$ is consistent with zero. \emph{Third row:} Same, for the case of a wCDM model. Here the CMB can not constrain these parameters simultaneously, hence the broad posteriors. The degeneracy between $\Omega_m - w$ is broken by adding either Type Ia SNe or our data. The HST $H_0$ prior was added to the $w$ run using our data.}
\label{fig:1d}
\end{center}
\end{figure}

\begin{figure*}
\begin{center}
\includegraphics[width=0.85\linewidth, angle=0]{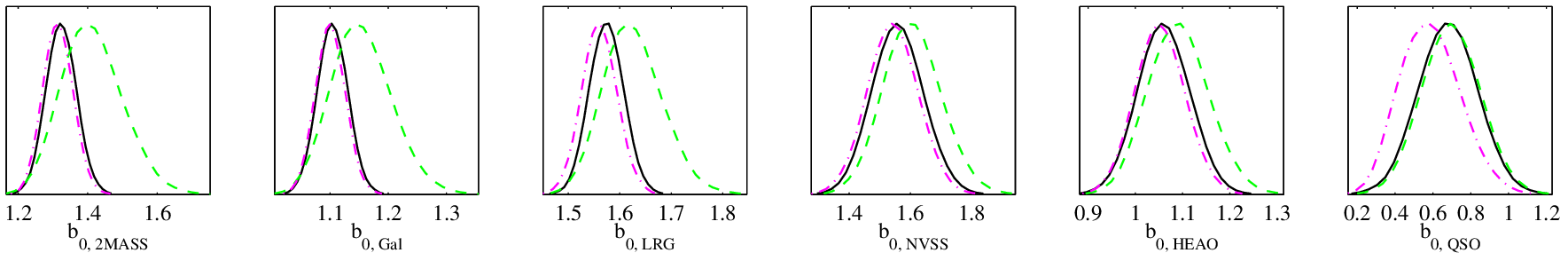}
\includegraphics[width=0.5\linewidth, angle=0]{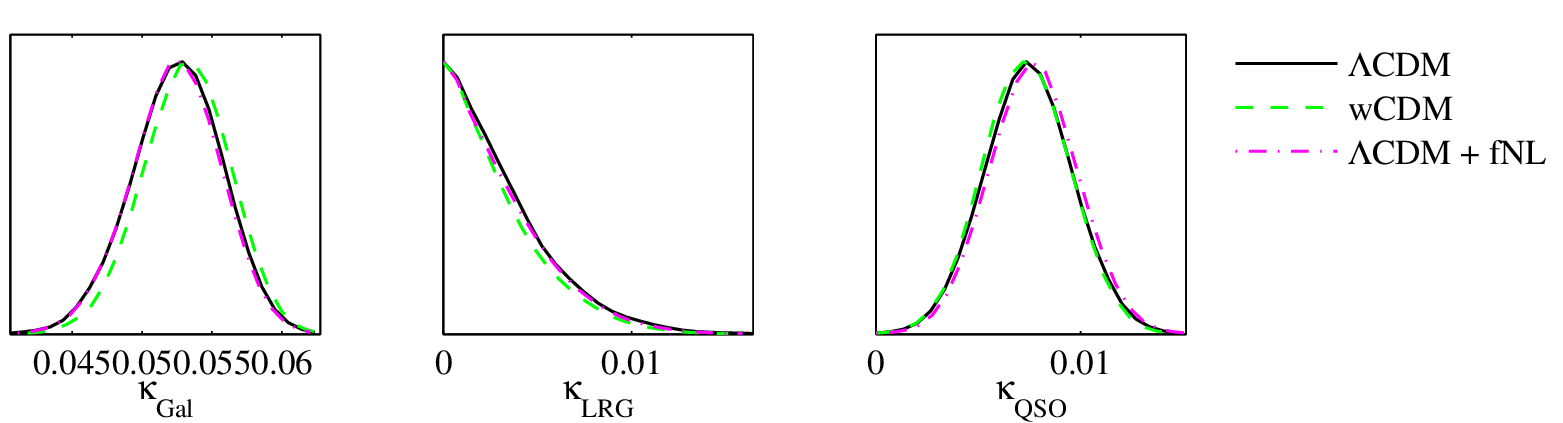}
\caption{Marginalized posterior probability distributions on a choice of nuisance parameters for different models. \emph{First row:} The bias parameters $b_0^i$ for each catalog.   \emph{Second row:} Same, for the stellar contamination parameters $\kappa_i$, which are introduced for the SDSS data only. The wCDM case includes the HST $H_0$ prior.}
\label{fig:1dnuise}
\end{center}
\end{figure*}

\section {Results}   \label{sec:results} 

We examine a range of models; starting from a simple flat \LCDM~model, we add  primordial non-Gaussianity of different types, or the dark energy equation of state $w$. The parameters which we use and their prior ranges are shown in Table~\ref{tab:params}. 

We study the cosmological consequences of these data via a Monte Carlo Markov Chain (MCMC) method using the publicly available \textsc{Cosmomc} code \cite{Lewis:2002a} and its extension to nested sampling \textsc{Multinest} \cite{2009MNRAS.398.1601F}, which replaces the Metropolis-Hastings with a nested sampling algorithm, achieving in the process higher accuracy, greater speed, and an estimation of the Bayesian evidence for each model. We ensured to always use high enough accuracy within \textsc{Multinest} and \textsc{Camb} to obtain numerical stability, also enforcing a sufficiently broad $k$ range in the integrations.

\begin {table}
\begin {center}
\begin{tabular}{| l | c | c | c |}
\hline
Catalog    &  $A \pm \sigma_A$   &  S/N  & expected \\
\hline
2MASS cut    & 1.32 $\pm$ 2.02   &  0.7 &  0.5 \\
SDSS gal DR8     & 1.29 $\pm$ 0.59   &  2.2 &  1.6 \\
\emph{SDSS LRG MegaZ}     & \emph{2.10 $\pm$ 0.84}   &  \emph{2.5} &  \emph{1.2} \\
\emph{SDSS LRG CMASS}     & \emph{1.10 $\pm$ 0.61}   &  \emph{1.8} &  \emph{1.6} \\
\hline
\emph{NVSS G12}       & \emph{1.21 $\pm$ 0.43}  & \emph{2.8} &  \emph{2.6} \\
\emph{NVSS corrected}      & \emph{1.47 $\pm$ 0.41}  & \emph{3.6} &  \emph{2.6} \\
\emph{NVSS raw}       & \emph{1.88 $\pm$ 0.51}  & \emph{3.7} &  \emph{2.6} \\
HEAO        & 1.37 $\pm$ 0.57   & 2.4  &  2.0 \\
SDSS QSO DR6     & 1.46 $\pm$ 0.60   & 2.4  &  1.7 \\
\hline
\textbf{TOTAL G12} & \textbf{1.38 $\pm$ 0.32} & \textbf{4.4 $\pm$ 0.4} &  $\sim$3.1 \\
\textbf{TOTAL, corr. NVSS} & \textbf{1.39 $\pm$ 0.31} & \textbf{4.5 $\pm$ 0.4} &  $\sim$3.1 \\
\textbf{TOTAL, raw NVSS} & \textbf{1.48 $\pm$ 0.33} & \textbf{4.6 $\pm$ 0.4} &  $\sim$3.1 \\
\hline
\end{tabular}
\caption{Single-catalog and total ISW significances when updating the LRGs and cleaning the NVSS data. A $\sim 0.4$ error on the full signal-to-noise ratio is added as in G12  to account for the typical changes resulting from different choices in the analysis. The result from the raw NVSS data shows a $\sim 1.7 \sigma$ excess compared with the \LCDM~expectations, increased from the measurement in G12 where we used the striping correction of the declination systematics. This increase is due to the higher level of measured CCF in the first few angular bins, that cause the best-fit template to become higher than all data points, due to the highly correlated error bars. The final result is only marginally affected. Further, the NVSS result obtained when correcting the observed systematics in r.a. and dec is also compatible within the errors, presenting a lower excess of $\sim 1.1 \sigma$ only.}
\label{tab:SN}
\end{center}
\end {table}

\subsection {\LCDM}
\label{sec:resultsLCDM}
\paragraph*{\textbf{ISW Significance}} We first consider a \LCDM~template with a single amplitude of the ISW $A$; as described in G08 and G12 this can be done by assuming the CCFs in each bin $i$ are written as $ w_i^{Tg} = A \, g_i$, where $g_i$ is the template prediction from the fiducial model. Then the best-fit $A$ can be calculated from the covariance matrix $\mathcal{C}$ and from the observed CCFs $\hat w_j^{Tg}$ by analytically maximizing the likelihood:
\be
A = \frac{\sum_{i,j = 1}^p \, \mathcal{C}_{ij}^{-1} \, g_i \, \hat w_j^{Tg}}{\sum_{i,j = 1}^p \, \mathcal{C}_{ij}^{-1} \, g_i \, g_j} \, ,
\ee
where in our case $p = 13$ if considering a single catalog, and $p = 13 \times 6 = 78$ for the full combined analysis. The variance is given by
\be
\sigma_A^2 = \left( \sum_{i,j = 1}^p \, \mathcal{C}_{ij}^{-1} \, g_i \, g_j \right)^{-1} \, .
\ee
In this way we recover the significances reported in Table~\ref{tab:SN}, which are consistent with our earlier work, and with the expected signal-to-noise ratio within the statistical errors, calculated according to Eq.~(6) in G12. The significance of the ISW detection in the CMASS LRGs is at $1.8 \sigma$, down from $2.5 \sigma$ which was found with the MegaZ data, as the measured CCF has decreased, in agreement with the results by Ref.~\cite{HM13}. The significance of the NVSS ISW with raw data is increased to $3.7 \sigma$, as the signal has increased. Given the highly correlated error bars and the different shape of the template to the data points, the best-fit curve happens to be higher than the data points. However the strong systematics that are present in NVSS will also increase the errors, and thus the actual significance is almost certainly lower than the one derived from the statistical error. The NVSS ISW significance is marginally lower at the $3.6 \sigma$ level if we correct the r.a. and dec systematics, and the results using different treatements are anyway fully compatible within the errors. The overall significance has remained nearly unchanged from G12 at $4.6 \sigma$, to which we add a $0.4 \sigma$ estimate of the systematic error as in G12.

\paragraph*{\textbf{Full MCMC}} We then fully test  a simple flat \LCDM~model using \textsc{Multinest}; we can see in the first row of Fig.~\ref{fig:1d} the resulting 1D marginalized posteriors on a choice of cosmological parameters. Adding our `fair' data to the WMAP7 CMB power spectrum yields results which agree within the errors, even though our data prefer a lower matter density $\Omega_m$: we find $ 0.21 < \Omega_m < 0.28 $ (95\% c.l.) when using our data and the CMB.
We show in Fig.~\ref{fig:1dnuise} the marginalized constraints on the most important nuisance parameters. We can see that about $5\%$ stellar contamination seems present in the main galaxy sample, which is reduced to $1\%$ in the quasars, while no evidence for residual contamination appears for the LRGs, in agreement with our positive assessment of the purity of this sample.

\begin{figure}
\begin{center}
\includegraphics[width=\linewidth, angle=0]{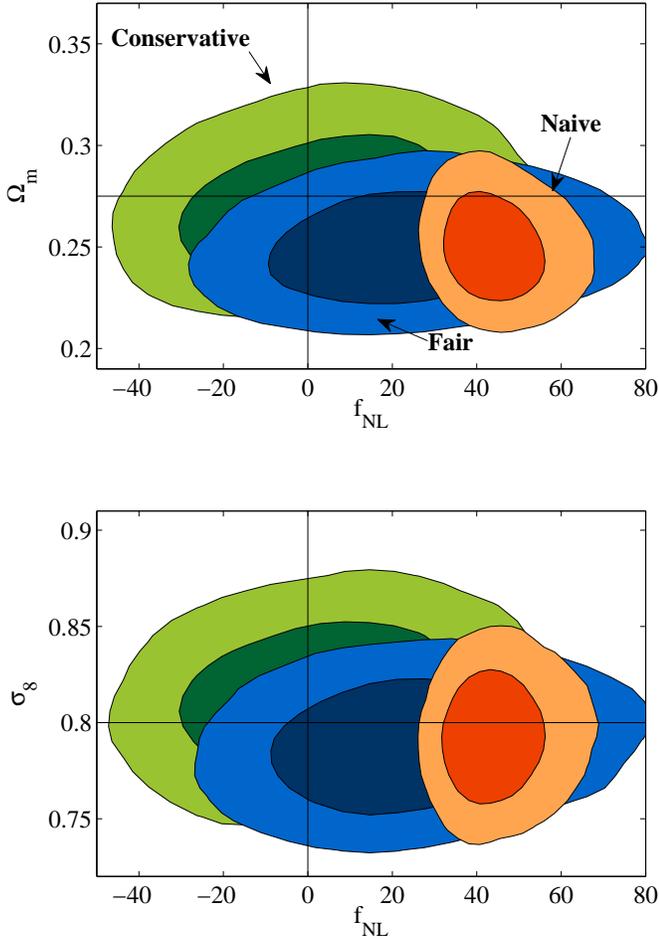}
\caption{Marginalized posterior probability distributions for $\fnl$ versus $\Omega_m, \sigma_8$ for different data sets. The na\"{i}ve result obtained using the complete data set (red contours, at $68$ and $95\%$ c.l.), which would suggest the presence of significant PNG, is not stable. When using only the most reliable parts of our compilation, we obtain the conservative result (green), which remains stable when adding back most of the data (blue), except the quasar and NVSS ACFs. As these are the least reliable data, as we discussed in our systematic section, we decide to discard them.}
\label{fig:fNL_2d}
\end{center}
\end{figure}

\subsection {Primordial Non-Gaussianity}
\paragraph*{\textbf{The ${{\boldsymbol{f}}_{\mathbf{NL}}}$ Model}} We next consider the local $\fnl$ model, which contains one extra parameter.
When  estimating the posterior likelihood of $\fnl$ from the whole `na\"ive' data set, including all auto-correlations, 
we find a strong positive detection: $ 31 < \fnl < 64 $ at 95\% c.l., as shown in 2D by the red contours of Fig.~\ref{fig:fNL_2d}.
This result is consistent with previous measurements that used similar data sets \cite{Xia:2011a}, but our uncertainty on $\fnl$ is reduced due to the fact that we:
(i) Effectively include all multipoles $l$ (for SDSS data, we include the $\kappa$ parameters to marginalize over the known low-$l$ issues);
(ii) Include fainter NVSS data (Ref.~\cite{Xia:2011a} applied a 10 mJy cut);
(iii) Include 33\% more area for the LRGs;
(iv) Include additional data sets.

To assess whether this result is robust, we study how it is changed when less reliable data are excluded. 
When we repeat this test with the `conservative' data, we find no detection:  $ -36 < \fnl < 45 $  at 95\% c.l., corresponding to the green contours of Fig.~\ref{fig:fNL_2d}.
When we examine the `fair' data,  we still see no hint for PNG (blue contours in Fig.~\ref{fig:fNL_2d}). 
By looking at both panels of Fig.~\ref{fig:fNL_2d}, we can see no evidence of strong degeneracies between $\fnl$ and $\Omega_m$ or $ \sigma_8$. The results are in agreement with the simplest Gaussian \LCDM~model: the marginalized 95\% c.l. interval from our `fair' sample is $ -15 < \fnl < 68 $.

\begin {table}
\begin {center}
\begin{tabular}{| c | c |}
\hline
    Data     &   $\fnl$ 95\% interval\\
\hline
 Conservative  &  $ -36 < \fnl < 45 $  \\
  Fair  &  $ -15 < \fnl < 68 $ \\
 Na\"{i}ve &  $ 31 < \fnl < 64 $   \\
\hline
   Conservative; with $\beta_{ij}$,  $\sigma_{\beta}^{\max} = \ln 2$ &   $ -36 < \fnl < 42 $   \\
   Conservative; with $\beta_{ij}$, $\sigma_{\beta}^{\max} = \ln 4$  &   $ -36 < \fnl < 42 $  \\
\hline
\hline
   LRG-LRG  &  $ -116 < \fnl < 91 $ \\
  NVSS-NVSS &  $ 140 < \fnl < 245 $ \\
   QSO-QSO  &  $ -13 < \fnl < 91 $ \\
\hline
\end{tabular}
\caption{Summary of the $\fnl$ constraints. In the top section we show the results from the combined analyses. The two runs with the nuisance parameters $\beta_{ij}$ have two different choices for their Gaussian priors. The runs with a single ACF below include marginalizations over one bias parameter and (for the SDSS catalogs) one stellar contamination parameter.}
\label{tab:fnl}
\end{center}
\end {table}

\begin{figure}
\begin{center}
\includegraphics[width=\linewidth, angle=0]{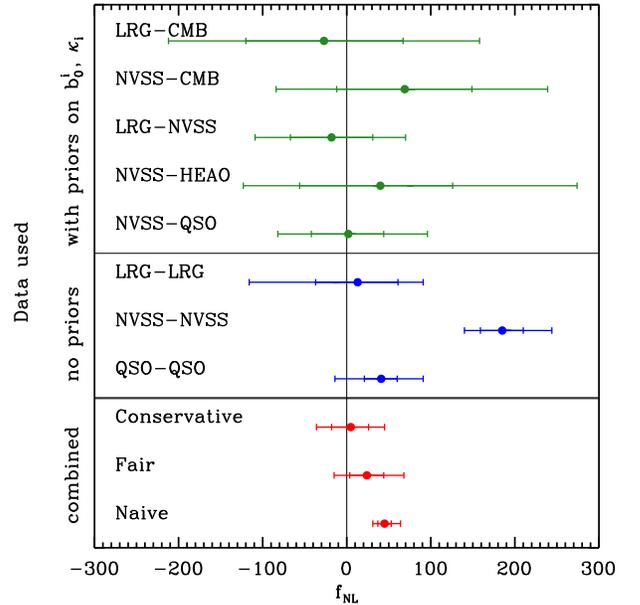}
\caption{Comparison of the marginalized posterior probability distribution on $\fnl$ using the parts of our data set giving the strongest contributions.
We show the results from single cross-correlation functions (top, green), auto-correlations (center, blue), and from combined sub-samples of the whole data set (bottom, red). The lines correspond to 68 and 95\% ranges, have been marginalized over the cosmological parameters, and include the WMAP7 CMB priors. The points represent the mean values of the posterior likelihoods. The results from single auto-correlation functions have also been marginalized over one bias parameter and one stellar contamination fraction (for the SDSS samples). The NVSS ACF result is inconsistent with the rest, but is discarded due to the high level of systematics.
To best present the relative constraining power of the cross-correlation measurements, we have placed priors on the bias and stellar contamination parameters, which significantly overstate the constraints these cross-correlation allow on their own. See the main text for more details.}
\label{fig:1D_fnl}
\end{center}
\end{figure}

We can see the corresponding marginalized 1D distributions in the second row of Fig.~\ref{fig:1d}; by comparing with the first row, we can also notice that best-fit regions of a selection of other cosmological parameters do not change when adding $\fnl$: this indicates overall low degeneracy between these parameters and $\fnl$.

We further tested for possible degeneracies between $\fnl$ and other parameters by examining the marginalized 2D likelihood contours, 
 particularly between $\fnl$, the stellar contamination $\kappa$ and the bias $b_0 $ for each sample. 
We found that the only parameter that presents some degeneracy with $\fnl$ is the quasar bias $b_0^{\mathrm{QSO}}$.
When considering the `na\"{i}ve' data set we found that, perhaps surprisingly, there is no significant degeneracy with the quasars' $\kappa$, while the degeneracy with $b_0 $ is more pronounced: by raising it to $b_0 \simeq 1$, values as low as $\fnl \simeq 30$ are allowed. When using the quasars' ACF however, the Gaussian limit is always excluded at $> 2 \sigma$. We found that a stronger degeneracy between $\kappa$ and $\fnl$ is present only when using the quasar ACF alone.
When using the `conservative' data set, we found that the $\fnl - b^{\mathrm{QSO}}_0$ degeneracy is significantly moderated.

We summarize the constraints on $\fnl$ in Table~\ref{tab:fnl} and in Fig.~\ref{fig:1D_fnl} for clarity. Here we compare the marginalized results obtained when using the most constraining parts of our data set. We can see once again that most results agree with Gaussian initial conditions, and with each other. When considering single auto-correlation functions, we  marginalize over cosmology including the WMAP CMB likelihood, and over one bias parameter and one stellar contamination fraction (for the data derived from SDSS).
To better interpret the cross-correlations on their own, we have assigned Gaussian priors on the relevant bias and stellar contamination parameters equal to the posteriors on these parameters obtained from the `fair' data. 
Applying these priors allows us to accurately portray the relative importance of each cross-correlation to our bottom-line results. Furthermore, we found that applying the bias prior to the auto-correlations would increase the precision of their $\fnl$ constraints by a factor of two. Accounting for this factor, the LRG auto-correlation is the best-constrained measurement that enters the `conservative' data set.
When using the LRG ACF only we recover a result consistent with the recent analysis by Ref.~\cite{Ross13fnl}, who found $-45<\fnl<195$ at 95\% using the spectroscopic sample of the CMASS LRGs, which contains $\sim 1/3$ of the photoz sample we use.

Notice that the factor $(b_1-1)$ within the bias correction $\Delta b$ is the leading contribution that determines the size of the $\fnl$ error bars. For this reason, the low-bias data from 2MASS, the SDSS main galaxies, and HEAO bring little information on $\fnl$. Also the external correlations of the quasars bring less contribution than it may be expected, since the quasar bias at low redshift is also low.
This explains why the strongest constraints come from NVSS, the LRGs and their external correlations.
For this reason, we have also checked the effect of the assumed NVSS bias evolution with one additional run where the evolution parameter $\gamma_{\mathrm{NVSS}}$ is let free, and we found no significant changes in the results.

\begin{figure}
\begin{center}
\includegraphics[width=\linewidth, angle=0]{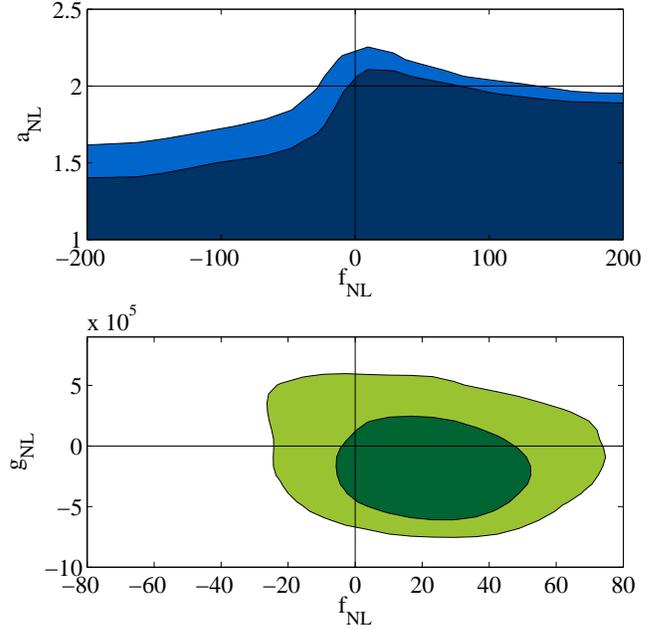}
\caption{Marginalized posterior probability distributions for extended PNG models. \emph{Top panel:} The parameter $\anl$ defines the scale-dependence of the bias and it is $\anl = 2$ in the local, scale-independent model. The two shaded contours represent the 95 and 99\% confidence regions. There is a vertical infinite degeneracy along the $\fnl = 2$ direction, which is only partially visible due to finite sampling resolution and smoothing. \emph{Bottom panel:} Marginalized constraints on the $\fnl - \gnl$ plane (68 and 95\% regions). As both parameters produce the same scale dependence of the bias, they are degenerate, but only partially, as the redshift dependence is different. Note that the $\gnl$ constraints are optimistic, as they assume the validity of the fitting formula by Ref.~\cite{2012JCAP...03..032S} for our data.}
\label{fig:aNL}
\end{center}
\end{figure}

\paragraph*{\textbf{The ${{\boldsymbol{a}}_{\mathbf{NL}}}$ Model}} We then extend our model to generalized PNG defined in~Eq.~(\ref{eq:anl}): in addition to $\fnl$, we thus allow for scale dependence of the bias of any slope $\anl$, which reduces to $\anl=2$ in the local, scale-independent case. We show our marginalized posterior likelihood distribution in the top panel of Fig.~\ref{fig:aNL}, where we can see that, in line with the lack of evidence for $\fnl$, there is no evidence for $\anl$ either. The full marginalized upper limit we find is $\anl < 1.8 $ at 95\%, but it must be born in mind that there is an infinite degeneracy along the direction $\fnl = 0$ by construction: thus, this result is strongly dependent on our adopted priors, rather than being a ``stand-alone measurement''. The correspondent bound on $\nfnl$ can be found using Eq.~(\ref{eq:an}).

\paragraph*{\textbf{The ${{\boldsymbol{g}}_{\mathbf{NL}}}$ Model}} We finally consider the $\gnl$ model. We shall here make the optimistic assumption that the fitting formula of Eq.~(\ref{eq:gNL}) is a reasonable approximation to the effect of $\gnl$, keeping in mind that this may not be accurate in all cases due to the low bias of our catalogs. Under this assumption we find $ -5.6 \cdot 10^5 < \gnl < 5.1 \cdot 10^5 $ (95\%) if assuming $\fnl=0$. However as shown by Refs.~\cite{2012JCAP...03..032S,2012MNRAS.425L..81R}, and as clear from Eq.~(\ref{eq:fnl101}), there is a degeneracy between $\fnl$ and $\gnl$, as both parameters produce a scale dependence of the bias of the same order $\sim k^{-2}$; the degeneracy is alleviated by the different redshift dependences. This is indeed what happens when we consider the complete model where both parameters are left free: we can see in the bottom panel of Fig.~\ref{fig:aNL} that the marginalized posterior partially presents this degeneracy, as demonstrated with $N$-body simulations by Ref.~\cite{2012MNRAS.425L..81R}. Also in this case the Gaussian model remains well within the 95\% region: the marginalized constraints on the two parameters are marginally degraded to $ -17 < \fnl < 63 $ and $ -5.9 \cdot 10^5 < \gnl < 4.7 \cdot 10^5 $ at 95\% respectively when they are both set free.

\subsection {Dark Energy}

\begin{figure}
\begin{center}
\includegraphics[width=\linewidth, angle=0]{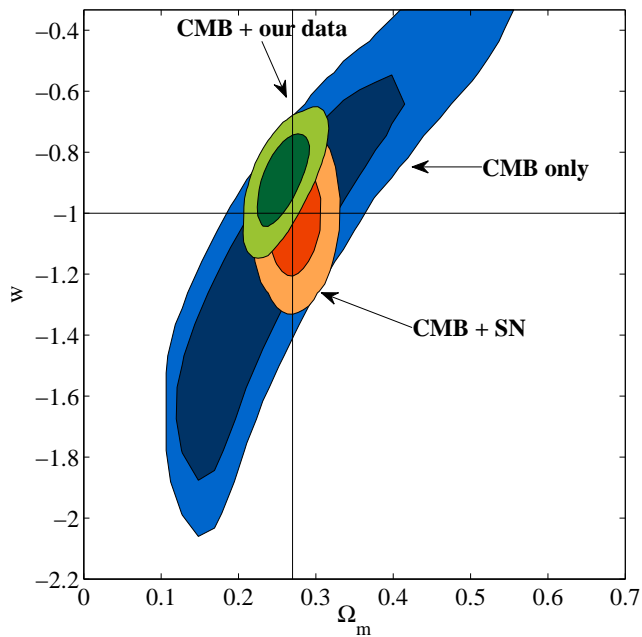}
\caption{Marginalized posterior probability distributions for $\Omega_m$ vs $w$ for different data sets: CMB only (blue), CMB + Type Ia Supernovae (red), and CMB + our data (green). The results are in agreement with the concordance \LCDM~model. The HST $H_0$ prior was added to the run using our data.}
\label{fig:w_2d}
\end{center}
\end{figure}

Without the ISW effect from the presence of dark energy, we would not be able to use the CMB cross-correlations to constrain PNG. In the above analysis, we jointly fitted the dark energy density and PNG to test the sensitivity of the PNG measurements to our assumptions about DE. 

\paragraph*{\textbf{The wCDM Model}} We next also consider models with Gaussian initial conditions, and use the ISW data to constrain the simplest dynamical dark energy of equation of state $w$. 
We present in the third row of Fig.~\ref{fig:1d} the marginalized 1D posterior probability distributions when considering this wCDM model. Here we can see that, as it is well known, the CMB temperature power spectrum alone can not break the $\Omega_m - w$ degeneracy, which is broken instead by either Type Ia Supernovae or our data. This can be seen more clearly in the 2D plot of Fig.~\ref{fig:w_2d}. Here we show that the results from our data and the CMB power spectrum reduce the parameter space significantly around the concordance values. Notice that we have added the HST prior \cite{2009ApJ...699..539R} on the Hubble constant $H_0 = 74.2 \pm 3.6$, since we found that otherwise significant degeneracies appear between $w$, the bias parameters, and $H_0$, due to compensating effects in the growth function $D(z)$.
When marginalizing over all other parameters, we find from our data $ -1.09 < w < -0.70 $ at 95\%. This confirms the overall agreement with the standard \LCDM~paradigm of the ISW measurements.

\begin {table*}
\begin {center}
\begin{tabular}{ | c | c | c | c | c | c|}
\hline
 Parameters   & Data     &    $\ln (\mathcal{Z})$ &  $\ln B = \Delta \ln (\mathcal{Z})$ & Odds & Interpretation \\
\hline
 \LCDM  & CMB + our                &    $ -3983.27 \pm 0.15 $  &    0    & --- & --- \\
 + $\fnl$  &     (`fair')          &    $ -3984.44 \pm 0.15 $  &   $ -1.17 \pm 0.21 $ & $ 1 : 3 $ &  Weak evidence against $\fnl$ \\
 + $\gnl$  &                       &    $ -3984.43 \pm 0.15 $  &   $ -1.16 \pm 0.21 $ & $ 1 : 3 $ & Weak evidence against $\gnl$ \\
 + $\fnl$ + $\gnl$  &              &    $ -3985.61 \pm 0.13 $  &   $ -2.34 \pm 0.20 $ & $ 1 : 10 $ & Weak evidence against $\fnl+\gnl$ \\
 wCDM  &           &    $ -3984.73 \pm 0.15 $  &   $ -1.46 \pm 0.21  $ & $ 1 : 4 $ & Weak evidence against $w$ \\
\hline
 \LCDM  & CMB alone    & $ -3751.37 \pm 0.08 $ &  $ 0 $ & --- & --- \\
 wCDM  &   & $ -3751.84 \pm 0.09 $ &   $ -0.47 \pm 0.12 $ & $ 5 : 8 $ & Inconclusive \\
\hline
 \LCDM  & CMB + SN     & $ -4017.15 \pm 0.09 $ & $ 0 $ & --- & --- \\
 wCDM  &    & $ -4019.10 \pm 0.09 $ &  $ -1.95 \pm 0.13 $ & $ 1 : 7 $ & Weak evidence against $w$ \\
\hline
\end{tabular}
\caption{Bayesian model selection. For each data set we compare models with extra parameters with the baseline \LCDM. We can see that our `fair' data in combination with the CMB provide weak evidence against models with one extra PNG parameter; the odds become worse against a more complex model with both $\fnl, \gnl$ present.}
\label{tab:evidence}
\end{center}
\end {table*}

\subsection{Model Selection}

Nested sampling allows model selection in addition to parameter estimation, thanks to the calculation of the Bayesian evidence factor for each model.
Briefly, if we assume a model $M$ of parameters $\Theta$, and we compare it with data $D$, Bayes' theorem states that the posterior probability distribution $\mathcal {P} $ on the parameters  is given by
\be
\mathcal {P}(\Theta) = \frac {\mathcal {L} (\Theta) \, \Pi (\Theta)}{\mathcal {Z} (M)} \, ,
\ee
where the prior is $\Pi (\Theta)= P (\Theta | M)$, the likelihood is $\mathcal {L}(\Theta) = P (D | \Theta, M)$, and the Bayesian evidence is $\mathcal{Z} (M) = P (D | M) $.
When interested in parameter estimation only, it is common to neglect the evidence $\mathcal{Z}$ and simply study the posteriors of arbitrary normalization for a given model. However here we will also compare different models (e.g. with and without PNG), and for this we will use the evidence. Model selection between two models $M, N$ can be performed by calculating the ratio of their evidences, also called Bayes factor: $B \equiv \mathcal{Z}_N / \mathcal{Z}_M$, so that $\ln B = \Delta \ln \mathcal{Z}$. This factor, which incorporates Occam's razor by penalizing models with unnecessary extra parameters, expresses the odds between the models given the data, and can be  qualitatively interpreted with the heuristic Jeffrey's scale, which states that the model selection is inconclusive if $|\ln B| < 1$, and that the evidence for one model is weak, moderate, or strong for $\ln B > 1$, $\ln B > 2.5$, $\ln B>5$ respectively \cite{2008ConPh..49...71T}. 

To test non-Gaussianity, we can compare the Bayesian evidence $\mathcal Z$ of the models with and without PNG. As shown in Table~\ref{tab:evidence}, the Bayes factor between these models is $ \Delta \ln \mathcal {Z}= -1.17 $ in the case of $\fnl$ and $ \Delta \ln \mathcal {Z}= -1.16 $ in the case of $\gnl$, which are both interpreted as weak evidence against these models according to Jeffrey's scale, as the two models have odds  $\sim 1 : 3$ compared with \LCDM~given our data \cite{2008ConPh..49...71T}.
The situation is more defined for the extended model with both $\fnl$ and $\gnl$: in this case we find  $ \Delta \ln \mathcal {Z}= -2.34 $, corresponding to odds of $1:10$. 

We also performed a Bayesian model selection for dark energy: as shown in Table~\ref{tab:evidence}, when using the CMB and our data this test is also weakly disfavoring the wCDM model:  $ \Delta \ln \mathcal {Z}= -1.46 $, which means  the odds are $1:4$ when comparing \LCDM~with wCDM. The result becomes more defined when combining CMB and data from Type Ia Supernovae: in this case $ \Delta \ln \mathcal {Z}= -1.95 $.

In general  it has to be kept in mind that the Bayesian selection is affected by the assumed choice of priors, as the prior normalization implies the evidence is decreased if the prior range is broader. We have however chosen a relatively narrow prior on $\fnl$ and $\gnl$ compared with existing constraints, meaning we are not unnecessarily penalizing the evidence for these PNG models.

\section {Conclusions} \label{sec:conclusion}

Scale-dependent bias has arisen as a key means of detecting primordial non-Gaussianity, but its detection can be severely compromised by a number of potential systematics.  These, if not accounted for, can introduce spurious large-scale structure and bias the inferred constraints on PNG. We have therefore conducted an analysis that carefully accounts for these systematic concerns and maximizes the information obtained from the most robust sources, in order to obtain the most precise to-date PNG measurements. 

In particular, large differences are observed when comparing the MegaZ and CMASS LRG samples. As shown in Ref.~\cite{Ross:2011a}, these appear to arise mainly from systematic relationships with foreground stars which, if unaccounted for, cause spurious large-scale power. Similarly, the NVSS auto-correlation is affected by severe systematic fluctuations in the source number density as a function of r.a. and declination. While correcting for these systematics is possible, such corrections are difficult to model in real space, and suppress the effect of PNG.  We have chosen here the most conservative approach of not applying any correction and discarding the NVSS ACF.
We have tested that that our results remain fully compatible if instead we apply a correction for the NVSS systematics.
  Finally, while quasar samples are potentially ideal for searching for primordial non-Gaussianity, we have confirmed earlier work \cite{Pullen_qso} showing that the quasar auto-correlation measurements are compromised by systematic errors.  In particular, there are correlations between the mean quasar density and a number of systematics including stellar density, galactic reddening and sky brightness.

While the existence of such systematics can be discouraging, they can often be corrected when they are understood.  In addition, cross-correlations between different data sets, which should generally contain different systematics, offer a much more robust means of addressing the question of primordial non-Gaussianity. 

One example of this can be seen in the analysis of cross-correlations with the CMB, previously analysed in G12.  Comparing the MegaZ and CMASS measurements, the CMB cross-correlation was seen to be much less affected than the auto-correlations.  While there were differences, these are partially accounted for by the larger sky coverage of CMASS; the extra area has associated cosmic variance and is potentially more subject to extinction systematics.  The signal decreased with respect to the G12 result, but the error bars are also smaller due to the larger DR8 sky coverage: the total signal-to-noise of this ISW detection alone is at the $1.8 \sigma$ level, in agreement with other recent measurements~\cite{HM13}. Similarly, using the raw NVSS data without subtracting the r.a. and declination dependent systematics changes its cross-correlation somewhat but not dramatically, actually raising the inferred signal by 20\%. However, the total ISW significance from the combined data is in the end nearly unchanged with respect to G12, as the increase in NVSS compensates the decrease in the LRGs. Also in this case, the results are consistent if we instead apply a correction to the NVSS systematics.

While measurements of cross-correlations are more robust, they require more information for their interpretation, in particular understanding how the surveys overlap in redshift.  We treat these uncertainties by marginalizing over nuisance parameters associated with the strength of the cross-corrrelation, so that it is its shape, rather than its amplitude, that constrains the PNG.  Effectively, the overlap is determined by the small scale cross-correlations, allowing us to use the large-scale measurements to constrain PNG.  \emph{Fundamentally, it is the lack of large-scale excess power in any of the cross-correlations which is responsible for our strong constraints.} The precise strength of the constraints will depend on our assumptions about the bias evolution, particularly for the deepest surveys; however, the bias models which we use are consistent with our physical understanding of the tracers and other observations. Further, we have applied a general bias model, one that effectively assumes only that the bias smoothly tends towards unity as the redshift decreases, to the NVSS sample and found no significant change in our results.

We have measured that the allowed amount of PNG is $ -36 < \fnl < 45 $ at 95\%. This bound was determined from our conservative analysis, for which we excluded all auto-correlations apart from the CMASS LRGs, and instead relied on the the cross-correlations between large-scale structure data sets and also CMB cross-correlations which are sensitive to scale-dependent bias through the ISW effect. We found the data {consistent with the simplest \LCDM~model in all cases}. This is significantly different than the positive signal detected if all possible auto-correlations are included, as the signal is driven up by the large-scale clustering in the quasar catalog.

Our measurements have also allowed us to constrain more general PNG models, as well as models of dark energy.  As yet, there are no indications of deviations from Gaussianity, or dynamic dark energy. 
The new Planck bispectrum constraints on PNG, released while this paper was being revised \cite{2013arXiv1303.5084P}, independently confirmed our findings.
Future large-scale structure measurements, such as will arise from DES, LSST and Euclid, may well determine whether the simplest inflationary models are correct.  However, the more sensitive the results, the more prone they are to potential systematics, making a careful comparison  and correlation between a range of different measurements indispensable.

\section*{Acknowledgments}

We thank Christian T. Byrnes, Marilena LoVerde, Sabino Matarrese, Hiranya Peiris, Cristiano Porciani and Jussi V\"{a}liviita for useful discussions.
TG acknowledges support from the Alexander von Humboldt Foundation, and from the Trans-Regional Collaborative Research Center TRR 33 -- ``The Dark Universe'' of the Deutsche Forschungsgemeinschaft (DFG). TG gratefully acknowledges the ICG Portsmouth for hospitality and for computing time on the Sciama facility.
 RC, RN, WP and AR gratefully acknowledge financial support from the STFC via the rolling grant ST/I001204/1. WP acknowledges funding from the European Research Council under the European Union's Seventh Framework Programme (FP/2007-2013) / ERC Grant Agreement 202686 -- MDEPUGS. AR acknowledges funding from The University of Portsmouth Research Infrastructure Funding Project.

Funding for SDSS-III has been provided by the Alfred P. Sloan Foundation, the Participating Institutions, the National Science Foundation and the US Department of Energy Office of Science. The SDSS-III web site is http://www.sdss3.org/.
SDSS-III is managed by the Astrophysical Research Consortium for the Participating Institutions of the SDSS-III Collaboration including the University of Arizona, the Brazilian Participation Group, Brookhaven National Laboratory, University of Cambridge, Carnegie Mellon University, University of Florida, the French Participation Group, the German Participation Group, Harvard University, the Instituto de Astrofisica de Canarias, the Michigan State/Notre Dame/JINA Participation Group, The Johns Hopkins University, Lawrence Berkeley National Laboratory, Max Planck Institute for Astrophysics, Max Planck Institute for Extraterrestrial Physics, New Mexico State University, New York University, Ohio State University, Pennsylvania State University, University of Portsmouth, Princeton University, the Spanish Participation Group, University of Tokyo, University of Utah, Vanderbilt University, University of Virginia, University of Washington and Yale University.

\appendix

\section{Fast Calculation of PNG Spectra}

Our analysis requires scanning over a large number of possible model parameters, which is potentially very computationally expensive if a CMB code such as \textsc{Camb} \cite{Lewis:2000a} must be rerun for every parameter choice.  Here we describe how we can make the calculation of the PNG effects more efficient, so that a range of bias choices and $\fnl$ values can be calculated using a single run of \textsc{Camb}.

\subsection {Gaussian Case}

In the Gaussian case, number density perturbations at a position $ \mathbf{x} $ and redshift $z$ are defined at linear level as
\be
\delta_g (\mathbf{x}, z)  = b_g(z) \, \delta (\mathbf{x}, z),
\ee
where $b_g(z)$ is the galaxy bias and $\delta (\mathbf{x}, z)$ are the underlying dark matter perturbations.
Projection onto the celestial sphere is achieved by summing over the past light cone, so that the 2D number density perturbation in a direction $\mathbf {\hat {n}}$ is
\be
\delta_g (\mathbf {\hat {n}}) = \int dz \, \varphi(z) \, b(z) \, \delta (\mathbf{x}, 0) \, D(z) \, ,
\ee
where $D(z)$ is the linear growth function normalized to unity today and $\varphi(z) $ is the normalized redshift distribution of the sources.

We can expand as usual these perturbations into harmonic space by means of a Fourier transform and an expansion in spherical harmonics $Y_{lm}$
\ba
\delta_g (\mathbf{x}, z) &=& \sum_{\mathbf k} \tilde \delta_g (\mathbf k, z) \, e^{i \mathbf {k} \cdot \mathbf {x}} = \\
&=& \sum_{\mathbf k} \tilde \delta_g (\mathbf k, z) \sum_{l, m} 4 \pi i^l j_l (kr) Y_{lm} (\Omega_{\mathbf {\hat k}}) Y_{lm}^{\ast} (\Omega_{\mathbf {\hat k}}) \nonumber \, ,
\ea
where $j_l$ are the spherical Bessel functions. We thus get the harmonic coefficients of the perturbations
\be
a_{lm} =  \sum_{\mathbf k} 4 \pi i^l \, \tilde \delta (\mathbf {k},z) \, Y_{lm}^{\ast} (\Omega_{\mathbf {\hat k}})
 \int dz \, \varphi(z) \, b_g(z) D(z) j_l(k r). \nonumber
\ee
We then introduce  for simplicity the quantity
\be
f_l (k) \equiv \int dz \, \varphi(z) \, b(z) D(z) j_l(k r),
\ee
so finally the projected angular power spectrum of the galaxy density perturbations can be written in a compact form in function of these quantities as
\be \label {eq:clgauss}
C_l^{gg} \equiv \langle a_{lm} a_{lm}^{\ast} \rangle = \frac {2}{\pi} \int dk \, k^2 \,  P(k) \, f_l^2(k) \, .
\ee

\subsection {Non-Gaussian Case}

When local non-Gaussianity of the $\fnl$ type is introduced, we know that at linear level the galactic bias will be corrected by
\be
\Delta b (k, z) = \frac {2 \, \fnl \, \delta_c \, [b_1(z) - 1]} {\alpha (k,z)} \, ,
\ee
where we have used the definition of $\alpha$ of Eq.~(\ref{eq:alpha}).
While redshift evolution of the bias is difficult to model, a general template can be written as a function of three parameters $b_0, b_{-1}, \gamma$ \cite{1998MNRAS.299...95M} as
\be
b_1(z) = b_{-1} + \frac{b_0 - b_{-1}}{D^{\gamma}(z)} \, .
\ee
We will here consider three simplified cases, introducing different templates for it, such as simply a constant
\be 
b_1 (z) = b_0 \, ,
\ee
or  
\be
b_1(z) = 1 + \frac {b_0 - 1}{D^{\gamma}(z)},
\ee
which satisfies the property of approaching unity in the future, and is therefore appropriate for flux-limited samples, and 
\be
b_1(z) = \frac {b_0} {D^{\gamma}(z)} \, ,
\ee
which instead tends to zero in the future, consistent with observations of quasars \cite{Pad09qso, White12qso}.

We can now generalize the 2D spectrum of Eq.~(\ref{eq:clgauss}) between two galaxy catalogs $i,j$ by noticing how the bias alterations due to PNG can be simply absorbed into the source function $f_l(k)$, which can be split in two: 
\ba \label{eq:clall}
 C^{g_i g_j}_l &=& \frac{2}{\pi} \int  d k \, k^2 \, P(k) \, \\
&\times& \left[ f_l^i (k) + \frac {g_l^i (k)} {\alpha(k,0)} \right] \, \, \left[ f_l^j (k) + \frac {g_l^j (k)} {\alpha(k,0)} \right] \,  ,  \nonumber
\ea
where we have absorbed the redshift dependence of $\alpha$ (a factor of $D$) into the integrand of $g_l^j$, and
\ba
f_l^i (k) &\equiv& \int dz \, \varphi_i(z) \, b_1^i(z)\,  D(z) \, j_l(k r)  \\
g_l^i (k) &\equiv& \int dz \, \varphi_i(z) \, \left\{2 \, \fnl \, \delta_c  \left[ b_1^i(z) - 1 \right] \right\} j_l(k r)  \nonumber \, .
\ea
In the integrand of $g_l^i$ we have simplified the factor of $D(z)$ from $\alpha$,  so that its original $D(z)$ has disappeared.

\subsection {Fast Decomposition of the Spectra}

It is useful to decompose the power spectra in a way which allows us to extract the bias and the $\fnl$ parameters outside of the integrals, so that the calculation is faster.

\subsubsection {Bias Parameterization $b_1(z) = 1 + \frac {b_0 - 1} {D^{\gamma}(z)}$}
To fix the ideas, let us assume for the redshift evolution of the bias that for a catalog $i$ it holds
\be
b_1^i(z) - 1 = \frac {b_0^i - 1} {D^{\gamma_i}(z)} \equiv \frac {b_{L0}^i} {D^{\gamma_i}(z)}.
\ee
We want now to express the power spectra using the parameter $b_{L0}^i$. 
The sources can be written explicitly as
\ba
f_l^i (k) &=& \int dz \, \varphi_i(z) \, D(z) \, j_l(kr)  \\
& & + \,  b_{L0}^i \int dz \, \frac{\varphi_i(z)}{D^{\gamma_i - 1}(z)} \,  j_l(kr) \nn \\
g_l^i(k) &=&   2 \, b_{L0}^i \, \fnl \, \delta_c \int dz \, \frac{\varphi_i(z)} {D^{\gamma_i}(z)} \,  j_l(kr)   \nn.
\ea

We can introduce here three source terms
\ba
f_l^{i,0} &\equiv& \int dz \, \varphi_i(z) \, D(z) \, j_l(kr)   \nn \\
f_l^{i,1} &\equiv& \int dz \, \frac{\varphi_i(z)}{D^{\gamma_i - 1}(z)} \,  j_l(kr)   \nn \\
g_l^{i,1} &\equiv&  2 \, \delta_c \int dz \, \frac{\varphi_i(z)} {D^{\gamma_i}(z)}  \,   j_l(kr)  \, ,
\ea
which are defined inside \textsc{Camb} in the file \texttt{equations.f90}, subroutine \texttt{output}.
The previous expressions can be simply rewritten as
\ba
f_l^i(k) &=& f_l^{i,0} (k) + b_{L0}^i \,  f_l^{i,1}(k) \nn \\
g_l^i(k) &=& \fnl \, b_{L0}^i \, g_l^{i,1}(k) \, ,
\ea
which can be done in \texttt{cmbmain.f90}, subroutine \texttt{calcScalCls}.
In this way the dependencies on both parameters $b_{L0}^i$ and $\fnl$ are factored out;
the power spectra can then be easily calculated using Eq.~(\ref{eq:clall}).

\subsubsection {Bias Parameterization $b_1(z) = \frac {b_0} {D^{\gamma}(z)}$}

In this case, $b^i_1 (z) - 1 =  \frac {b^i_{0}} {D^{\gamma_i}(z)} - 1 =  \frac {b^i_{L0} + 1} {D^{\gamma_i}(z)} - 1 $, so that we can write (notice here we use $b_0$ as nuisance parameter):
\ba
f^i_l (k) &=&  b^i_{0} \int dz \, \frac{\varphi_i(z)}{D^{\gamma_i - 1}(z)} \,  j_l(kr) \nn \\
g^i_l(k) &=&  -2 \,  \fnl \, \delta_c \int dz \, \varphi_i(z) \,  j_l(kr)   \nn \\
&~& + \, 2 \, b^i_{0} \, \fnl \, \delta_c \int dz \, \frac{\varphi_i(z)}{D^{\gamma_i}(z)} \,   j_l(kr)   \, .
\ea

We can define then the source terms
\ba
f_l^{i,1} &\equiv&  \int dz \, \frac{\varphi_i(z)}{D^{\gamma_i - 1}(z)} \,  j_l(kr) \nn \\
g_l^{i,0} &\equiv& - \delta_c \int dz \, \varphi_i(z) \,  j_l(kr)    \nn \\
g_l^{i,1} &\equiv&   2  \delta_c \int dz \, \frac{\varphi_i(z)}{D^{\gamma_i}(z)} \,   j_l(kr)   \, ,
\ea
so the previous expressions can be rewritten as
\ba
f^i_l(k) &=&  b^i_{0} \, f_l^{i,1}(k) \nn \\
g^i_l(k) &=& \fnl \, g_l^{i,0}(k) + \fnl \, b^i_{0} \, g_l^{i,1}(k) \, .
\ea

The constant bias case $b_1(z) = b_0$ can be also derived from these expressions.

\bibliography{ms}

\label{lastpage}

\end{document}